\documentclass[pdflatex,sn-basic]{sn-jnl}

\usepackage{graphicx}
\usepackage{amsmath,amssymb,amsfonts,mathtools,bm}
\usepackage{booktabs}
\usepackage{array,tabularx}
\usepackage{multirow}
\usepackage{enumitem}
\usepackage{microtype}

\newcommand{\ARI}{\mathrm{ARI}}

\newcommand{\op}{\mathrm{op}}

\newcommand{\tr}{\mathrm{tr}}

\begin{document}

\title[Regularized high-dimensional trimmed FKM and RKM]{Regularized trimmed Factorial and Reduced K-means for contaminated high-dimensional data: separating geometry, search and selection}

\author[1]{\fnm{Cinzia} \sur{Di Nuzzo}}\email{cinzia.dinuzzo@unict.it}

\author*[2]{\fnm{Matteo} \sur{Farn\`e}}\email{matteo.farne@unibo.it}

\affil[1]{\orgdiv{Department of Economics and Business}, \orgname{University of Catania}, \orgaddress{\street{Corso Italia 55}, \city{Catania}, \postcode{95129}, \country{Italy}}}

\affil*[2]{\orgdiv{Department of Statistical Sciences ``Paolo Fortunati''}, \orgname{University of Bologna}, \orgaddress{\street{Via delle Belle Arti 41}, \city{Bologna}, \postcode{40126}, \country{Italy}}}

\abstract{Factorial and Reduced K-means estimate a partition and a low-dimensional clustering subspace jointly, which makes them attractive when the variables that separate groups are not those that dominate total variance. On contaminated high-dimensional data they face three linked problems: trimming can make the scatter matrices singular, the criterion has many local optima, and the rank, number of groups and trimming proportion are unknown and interact. We propose REG-HD-TFKM and REG-HD-TRKM, regularized trimmed versions of the two methods, together with a selection procedure that handles these problems as separate decisions. A ridge-stabilized spectral step keeps every fit well defined; heterogeneous multistart search explores alternative solutions; method-specific agreement and reproducibility scores choose among them; and trimming is applied only when a radial diagnostic or the path over trimming levels supports it. A simulation study with 41 design cells shows why the separation matters. Two representations of euro-area bank balance sheets show that similar partitions can rest on nearly orthogonal subspaces, and that the same selected structure can describe different retained samples.}



\pacs[ORCID]{Cinzia Di Nuzzo: 0000-0003-2062-8930; Matteo Farn\`e: 0000-0002-2403-6599}

\maketitle

\section{Introduction}\label{sec:introduction}

Consider several hundred banks described by several hundred balance-sheet items. Most of the variation in such data comes from size and from items that every bank reports, whereas the differences between business models may sit in a handful of market, derivative and funding positions. A clustering method that first compresses the data by principal components keeps the directions of largest variance and can discard exactly the directions that separate the groups. Factorial K-means (FKM) and Reduced K-means (RKM) avoid this by estimating the partition and the clustering subspace together \citep{desoete1994kmeans,vichi2001factorial,timmerman2010factorial}. FKM asks for groups that are compact after projection; RKM also asks that the projection lose little of the data.

Real data of this kind are rarely clean. A few institutions with unusual activity can pull the subspace towards themselves, and the partition, which is assessed in that subspace, moves with it. Trimmed versions of FKM and RKM (TFKM and TRKM, respectively) handle this by leaving a fraction $\alpha$ of the observations out of the fit \citep{greco2020impartial,farne2025trimmed}. Three problems appear when these methods are pushed to dimensions comparable with, or larger than, the retained sample. First, after trimming the scatter matrices that drive the subspace update become singular, so the classical spectral step is no longer defined. Second, the criterion is non-convex, and different starting points end in different partitions, retained sets and subspaces. Third, the rank $r$ of the clustering subspace, the number of groups $g$ and the trimming proportion $\alpha$ are unknown, and each changes the geometry in which the others are assessed.

The central observation of this paper is that different causes of failure look the same in a final summary. A poor Adjusted Rand Index \citep[ARI;][]{hubert1985comparing} can mean that the data do not support the grouping in any low-dimensional projection, that the multistart search never reached a good solution, or that a good solution was reached but lost the comparison with competing fits. Conversely, a high ARI does not mean that the rank, number of groups or trimmed set have been recovered. Which of these happened matters, because each calls for a different remedy: more data, more or better starts, or a better selection rule.

We therefore propose REG-HD-TFKM and REG-HD-TRKM, regularized (REG) versions of TFKM and TRKM for high-dimensional (HD) data, built so that each decision is taken separately and can be inspected. At a fixed $(r,g,\alpha)$, a scale-aware ridge makes the spectral step well defined after trimming, and the fit alternates between subspace update, reassignment and impartial trimming. Each candidate is fitted from 100 starts drawn from four different initialization families. A method-specific score then chooses which terminal fit represents the candidate: for TFKM, agreement across initialization families; for TRKM, whether centroids rebuilt on part of the retained data reproduce the candidate's labels on the rest. The chosen fits are then compared across $(r,g)$ and followed along the grid of trimming levels. Positive trimming is adopted only when a radial diagnostic shared by both procedures detects an outer population, or when the path supplies method-specific evidence for it; the structure is then simplified only when the simpler state keeps essentially the same clustering.

A simulation study with 41 design cells and 30 replications per cell evaluates the complete procedures, treating the trimming decision, the structure $(r,g)$ and the partition as separate targets. In this design, the trimming decision is rarely wrong and diffuse leverage is always detected and trimmed at the nominal level, whereas the structure is the harder target, especially under diffuse leverage. A known-parameter benchmark shows that, given $(r,g,\alpha)$, the fitting cores are about as accurate as tandem methods on robust principal components for clean and bridge-contaminated data, but less accurate under strong diffuse leverage, where the multistart search does not always reach the generating state. A weak-signal experiment then shows where accuracy is lost: a truth-assisted projection classifies every observation correctly, yet in one regime no start reaches a good solution, and in another a good solution is reached but not always selected.

An application to euro-area bank balance sheets illustrates the distinction between partition, subspace and retained sample. Under ratios to total assets, the two procedures produce partitions with ARI $0.83$ on the commonly retained banks, but their clustering subspaces are almost orthogonal. Under log-level reporting variables, both select one dimension and two groups, yet they describe different retained populations.

The paper makes three contributions. (i) It provides REG-HD-TFKM and REG-HD-TRKM, which remain well defined when the retained scatter matrices are singular, together with a complete and reproducible rule for choosing $(r,g,\alpha)$. (ii) It separates the evidence used to choose a local solution from the evidence used to compare structures and to decide whether to trim. (iii) It proposes an evaluation design that distinguishes statistical geometry, access to a good solution and selection of that solution, and reports structure, trimming and partition as separate targets.

The remainder of the paper is organized as follows. Section~\ref{sec:literature} reviews related work. Section~\ref{sec:method} develops the procedures, Section~\ref{sec:simulation} reports the simulation study and Section~\ref{sec:application} the banking application. Section~\ref{sec:conclusion} discusses limitations and extensions. Appendix~\ref{sec:supp-implementation} gives the implementation details needed to reproduce the procedures exactly, and Appendix~\ref{sec:supp-sim} defines the truth-assisted reference quantities and reports additional simulation diagnostics.

\section{Related work}\label{sec:literature}

\paragraph{Joint clustering and dimension reduction.}
In a tandem analysis the reduced representation is chosen before clustering, without reference to the partition that will be estimated \citep{arabie1994cluster,desoete1994kmeans}. Principal component analysis keeps directions of large total variance, which need not be those that separate the groups. FKM and RKM estimate the representation and the partition jointly \citep{desoete1994kmeans,vichi2001factorial}. FKM minimizes within-group dispersion after projection, whereas RKM also accounts for the variation left outside the subspace, so the two criteria can prefer different geometries \citep{timmerman2010factorial}. Related joint approaches build each component from a disjoint subset of variables \citep{vichi2009clustering} or cluster functional data in a reduced subspace \citep{gattone2012clustering}. Factor discriminant K-means uses discriminant information more directly \citep{rocci2011new}. Strong consistency is known for K-means and, under suitable assumptions, for RKM and FKM \citep{pollard1981strong,terada2014strong,terada2015strong}; these results concern untrimmed criteria with fixed rank and number of groups. Model-based alternatives include HDDC \citep{bouveyron2007high}, Fisher-EM \citep{bouveyron2012fisher} and the broader family of parsimonious mixtures \citep{bouveyron2014model}; they rely on distributional assumptions, whereas FKM and RKM keep a centroid-based least-squares interpretation.

\paragraph{Robust clustering and trimming.}
Impartial trimming lets a fixed fraction of observations stay outside the fitted criterion; trimmed K-means introduced the idea \citep{cuesta1997trimmed}, and TCLUST combines it with constraints that prevent degenerate scatter estimates \citep{garciaescudero2008general,fritz2012tclust}. The trimming fraction is itself a consequential choice that changes both the groups and the set of representative observations \citep{torti2019trimming,garciaescudero2024review}, and recent work selects it jointly with the number of clusters \citep{garciaescudero2026choice}. Robustness can also be placed in the representation: spatial-sign PCA \citep{locantore1999robust} and ROBPCA \citep{hubert2005robpca} estimate robust principal components on which ordinary clustering can be run, and sparse robust K-means combines variable weighting with robustification \citep{witten2010framework,brodinova2019robust}. Trimming has been extended to more complex objects such as functional directional data \citep{alvarez2022robust}. Closest to our setting are trimmed reduced and factorial K-means \citep{greco2020impartial}, robust mixtures of factor analyzers \citep{yang2017robust} and trimmed FKM with iterative trimming in the reduced space \citep{farne2025trimmed}. These methods establish the value of robust joint subspace clustering, but they assume well-conditioned scatter matrices and do not address how to choose $r$, $g$ and $\alpha$ together.

\paragraph{High-dimensional clustering, structural selection and multistart search.}
High-dimensional clustering needs some restriction on where the group information lies. Sparse clustering assumes that it is carried by a subset of variables \citep{witten2010framework}; projected clustering allows different variables for different groups \citep{aggarwal1999projected,kriegel2009survey}; union-of-subspaces methods place groups near different subspaces \citep{vidal2011subspace,elhamifar2013sparse,liu2013robust}. FKM and RKM instead place all centroids in one common subspace. Random projections can diversify partitions in an ensemble \citep{mori2026random}; here they are used only to diversify starting points, while the fitted subspace is always given by the spectral step. For the number of groups, criteria based on the gap statistic \citep{tibshirani2001gap}, distortion \citep{sugar2003clusters}, prediction strength \citep{tibshirani2005cluster} or discriminant scores \citep{coraggio2023selecting} target different notions of separation and stability. Robust joint subspace clustering adds two further structural choices, $\alpha$ and $r$, and changing any of the three changes the geometry used to judge the others. Stability-based validation, finally, presupposes that the search has produced a good solution to recognize. The three possible failures listed in Section~\ref{sec:introduction} are rarely separated in empirical evaluations; the procedures and the simulation design below are built around that separation.

\section{The procedures}\label{sec:method}

A single objective value cannot answer all the questions raised in Section~\ref{sec:introduction}. For a proposed structure $(r,g,\alpha)$ we first need a fit that is well defined. We then need to decide which of the many local solutions should represent that structure, and only after that can structures be compared across ranks, group numbers and trimming levels. The order matters, because changing the retained observations changes the scatter matrices, changing the rank changes the distances used for assignment, and changing the number of groups changes the between-group signal.

REG-HD-TFKM and REG-HD-TRKM are therefore organized as a sequence of decisions, summarized in Table~\ref{tab:workflow}. Sections~\ref{sec:method-criteria}--\ref{sec:method-fit} describe the fit at fixed $(r,g,\alpha)$; Section~\ref{sec:method-multistart} the multistart search and the choice of one solution per candidate; Sections~\ref{sec:method-path}--\ref{sec:method-backward} the structural comparison, the trimming decision and the simplification step; and Section~\ref{sec:method-output} what the procedures return. All numerical constants were set while the procedures were developed on pilot designs with different seeds; none was tuned on the results reported below, and the same values are used in every analysis. Tie-breaking conventions and other details needed for exact reproduction are given in Appendix~\ref{sec:supp-implementation}.

\begin{table}[htbp]
\centering
\caption{The stages of REG-HD-TFKM and REG-HD-TRKM, with the question each stage answers.}
\label{tab:workflow}
\small
\begin{tabularx}{\linewidth}{@{}p{0.17\linewidth}p{0.33\linewidth}X@{}}
\toprule
Stage & Question & Device \\
\midrule
Fixed-parameter fit & Can a subspace--partition fit be computed for a given $(r,g,\alpha)$ when the retained scatter matrices are singular? & Ridge-stabilized generalized eigenproblems, alternated with reassignment and impartial trimming. \\
Multistart search & Does the solution reflect the data or only one starting point? & 100 deterministic starts from four initialization families. \\
Choice of solution & Which local solution should represent the candidate? & Agreement across initialization families for TFKM; reproducibility of labels in common robust views for TRKM. \\
Structural path & Which structure is preferred at each trimming level? & Comparison over 15 $(r,g)$ pairs at each of 10 trimming levels. \\
Trimming and simplification & Is positive trimming supported, and can the structure be simplified without moving to a different solution? & Radial diagnostic shared by both methods, then method-specific path evidence; backward moves accepted only when the clustering is preserved. \\
Output & What has been identified? & Structure, trimming decision and level, ambiguity flag and, when defined, a terminal partition and trimming mask. \\
\bottomrule
\end{tabularx}
\end{table}

\subsection{Two criteria at a fixed structure}\label{sec:method-criteria}

Let $X=(x_1,\ldots,x_n)'\in\mathbb R^{n\times p}$, let $U\in\mathbb R^{p\times r}$ have orthonormal columns with projector $P_U=UU'$, let $z\in\{1,\ldots,g\}^n$ be a label vector with membership matrix $G\in\{0,1\}^{n\times g}$, and let $M\in\mathbb R^{g\times r}$ contain the projected centroids. The rank $r$ is the dimension of the clustering subspace, $g$ the number of groups and $\alpha$ the fraction of observations left out of the fit. For a retained set $I$ with $|I|=h=n-\lfloor n\alpha\rfloor$, let $D_I$ be the diagonal indicator matrix of $I$ and $X_c$ the data centered on $I$. The TFKM criterion measures within-group dispersion inside the subspace,
\[
L_F(I,G,M,U)=\left\|D_I^{1/2}(X_cU-GM)\right\|_F^2,
\]
whereas the TRKM criterion measures how well group centroids constrained to the subspace reconstruct the retained observations,
\[
L_R(I,G,M,U)=\left\|D_I^{1/2}(X_c-GMU')\right\|_F^2.
\]
For orthonormal $U$,
\[
L_R=\left\|D_I^{1/2}X_c(I_p-UU')\right\|_F^2+L_F.
\]
This decomposition explains why the two methods can prefer different subspaces even when their partitions are similar: TFKM only asks that groups be compact after projection, while TRKM also penalizes the variation that the projection discards.

Trimming is carried out in the fitted subspace, inside the same alternating scheme, rather than as a preliminary screening step. Given a projector $P$ and the retained centroids $\bar x_k$, the means of the retained observations assigned to group $k$, each observation is assigned to the nearest projected centroid using
\[
d_{ik}^2(P)=\|P(x_i-\bar x_k)\|^2,
\]
and the $\lfloor n\alpha\rfloor$ observations with the largest nearest-centroid distances are excluded from the next retained set. The proportion $\alpha$ is therefore the share of the sample allowed to stay outside the fit, not the level of an outlier test.

\subsection{A regularized subspace step}\label{sec:method-ridge}

The classical spectral update breaks down precisely in the regimes of interest. After trimming, the retained total and within-group scatter matrices can be rank deficient even when the full data matrix is not, and the generalized eigenproblem is then ill posed. For a retained state $s=(I,z)$, let
\[
H_s=G_s(G_s'G_s)^{-1}G_s'
\]
and, after centering on the retained observations,
\[
S_T=X_c'D_IX_c,\qquad
S_W=X_c'(D_I-H_s)X_c,\qquad
S_B=X_c'H_sX_c=S_T-S_W.
\]
We add small ridges that scale with the average nonzero eigenvalue of each matrix,
\begin{equation}
\lambda_W=\tau_W\frac{\operatorname{tr}(S_W)}{\max\{\operatorname{rank}(S_W),1\}},
\qquad
\lambda_T=\tau_T\frac{\operatorname{tr}(S_T)}{\max\{\operatorname{rank}(S_T),1\}},
\label{eq:ridge}
\end{equation}
with $\tau_W=10^{-5}$ and $\tau_T=10^{-3}$ throughout. These constants only stabilize the computation and are kept fixed.

The subspace update replaces the classical spectral step by a ridge-stabilized generalized eigenproblem. REG-HD-TFKM takes as subspace the span of the $r$ generalized eigenvectors $u$ with the smallest generalized eigenvalues $\eta$ of
\begin{equation}
(S_W+\lambda_WI_p)u=\eta(S_T+\lambda_TI_p)u,
\label{eq:tfkm-eigen}
\end{equation}
and REG-HD-TRKM the span of the $r$ generalized eigenvectors with the largest generalized eigenvalues of
\begin{equation}
S_Bu=\eta(S_T+\lambda_TI_p)u.
\label{eq:trkm-eigen}
\end{equation}
The generalized eigenvalue is $\eta=u'(S_W+\lambda_WI_p)u/u'(S_T+\lambda_TI_p)u$ for TFKM, the within-group share of the retained variation along $u$, and $\eta=u'S_Bu/u'(S_T+\lambda_TI_p)u$ for TRKM, the between-group share. Small values for TFKM and large values for TRKM therefore select directions in which the groups are well separated. When $p\ge h$, $S_T$ has a null space, on which $\eta=\lambda_W/\lambda_T$ for TFKM; but $S_W$, of rank at most $h-g$, then also vanishes on $g-1$ directions of the row space of the retained data, where $\eta$ is close to zero. Since $r<g$, the TFKM directions are taken from these row-space directions and never from the null space of $S_T$.
A thin QR factorization \citep{golub2013matrix} turns the selected span into a Euclidean-orthonormal basis $U$, because assignment and trimming use the Euclidean projector $P_U=UU'$ rather than the raw generalized eigenvectors. These updates are ratio-normalized, ridge-stabilized versions of the classical steps rather than exact block minimizations of $L_F$ and $L_R$, and they remain defined when the scatter matrices are singular. At a fixed retained state the TFKM update minimizes $\tr\{V'(S_W+\lambda_WI_p)V\}$ and the TRKM update maximizes $\tr(V'S_BV)$, both subject to $V'(S_T+\lambda_TI_p)V=I_r$; the selected subspace is unique whenever the $r$th and $(r+1)$st generalized eigenvalues differ.

\subsection{One fit at fixed $(r,g,\alpha)$}\label{sec:method-fit}

Both procedures use the same alternating algorithm. Starting from an initial retained partition, each iteration
\begin{enumerate}[leftmargin=*]
\item centers the data on the currently retained observations;
\item recomputes $S_W$, $S_T$ and $S_B$;
\item updates the subspace through Equation~\eqref{eq:tfkm-eigen} or~\eqref{eq:trkm-eigen};
\item recomputes the retained projected centroids;
\item reassigns every observation to its nearest projected centroid;
\item trims the $\lfloor n\alpha\rfloor$ largest residuals.
\end{enumerate}
Because the subspace update is a ratio-normalized step rather than an exact minimization of $L_F$ or $L_R$, the criterion need not decrease monotonically, and convergence is not guaranteed. The algorithm stops when neither the assignment vector nor the trimming mask changes, or after 50 iterations. Because this map is run 15,000 times per data set and method (150 candidates, 100 starts each), it is made fully deterministic, including the repair of empty groups and the handling of ties (Appendix~\ref{sec:supp-fit}). A start is discarded only when the fit cannot be represented numerically. Reaching the iteration cap is recorded but does not invalidate a fit.

Each terminal fit also carries its retained latent residual
\[
J_{\rm trim}(U,z,I)=\sum_{i\in I}\min_{1\le k\le g}\|U'(x_i-\bar x_k)\|^2,
\]
which is used to break ties but not to choose among local solutions.

\subsection{Exploring and choosing local solutions}\label{sec:method-multistart}

\paragraph{Why multistart search.}
The alternating map is non-convex: two reasonable starts can converge to different partitions, retained sets and subspaces. A procedure can therefore fail in two distinct ways: the good solution is never reached, or it is reached but not recognized. Section~\ref{sec:sim-weak} measures both.

Each candidate $(r,g,\alpha)$ is fitted from 100 deterministic starts drawn from four initialization families. Variables are first standardized robustly by their median and median absolute deviation (MAD),
\[
X^{\rm rob}_{ij}=\frac{X_{ij}-\operatorname{med}_{i'}X_{i'j}}
{\operatorname{med}_{i'}|X_{i'j}-\operatorname{med}_{\ell}X_{\ell j}|+\varepsilon_s},
\]
where $\varepsilon_s$ is double-precision machine epsilon. The families are:
\begin{enumerate}[leftmargin=*]
\item \textbf{WINSOR--PCA--TKM}: trimmed K-means on principal components of $X^{\rm rob}$ winsorized to $[-4,4]$;
\item \textbf{SIGN--PCA--TKM}: trimmed K-means on $X^{\rm rob}$ projected onto directions obtained from the rowwise spatial signs;
\item \textbf{TKM--ORIGINAL}: trimmed K-means on all original variables after robust standardization;
\item \textbf{RP--TKM}: trimmed K-means after a Gaussian random projection.
\end{enumerate}
For the robust starting directions we use spatial-sign PCA \citep{locantore1999robust} rather than ROBPCA \citep{hubert2005robpca} because it has no tuning constant: ROBPCA would add a coverage parameter alongside the trimming levels being compared. With the true parameters supplied, trimmed K-means is almost equally accurate on either set of scores (mean ARI 0.976 and 0.974; Table~\ref{tab:competitors}).
The families present the same alternating fit with different starting geometries. Their purpose is to make a good solution easier to reach and to reveal solutions that appear only under one kind of start. TFKM allocates $(50,35,10,5)$ starts to the four families and TRKM $(40,30,15,15)$; the allocation affects the search only and plays no role in the choice of solution. Given the data and a candidate seed, the whole pool is reproducible (seeding rules in Appendix~\ref{sec:supp-fit}).

A large pool solves the exploration problem but leaves open which terminal fit should represent the candidate. We base this choice on evidence suited to each method rather than on the smallest residual. For TFKM the evidence is agreement across families: a convincing partition should reappear when the search starts from different robust geometries. For TRKM it is reproducibility: the labels of a fit should be recoverable from centroids rebuilt on part of the retained data. The two resulting scores, HIER4 and RV--CV, are internal components of the procedures and are not proposed as general cluster-validity indices.

\paragraph{TFKM: hierarchical agreement across the four families (HIER4).}
The name HIER4 refers to the hierarchical comparison over the four initialization families.
For two valid terminal fits $a$ and $b$ with trimmed sets $T_a$ and $T_b$, let $I_{ab}=T_a^c\cap T_b^c$ be the observations retained by both, where $T^c$ denotes the complement of $T$ in $\{1,\ldots,n\}$. Agreement combines the ARI on mutually retained observations,
\[
A_{ab}=\ARI\!\left(z_a|I_{ab},z_b|I_{ab}\right),
\]
with the Jaccard agreement of the trimmed sets,
\[
J_{ab}=\frac{|T_a\cap T_b|}{|T_a\cup T_b|},
\qquad J_{ab}=1\text{ if }T_a\cup T_b=\varnothing,
\]
through
\[
H_{ab}=A_{ab}+0.2J_{ab}.
\]
The weight $0.2$ lets agreement on trimming count without outweighing agreement on labels. To prevent a family with a larger budget from winning simply by contributing more near-duplicate fits, the choice is hierarchical. Within family $f$ with $m_f$ valid fits, the representative maximizes
\[
h_j^{(f)}=\frac1{m_f}\sum_{\ell=1}^{m_f}H_{j\ell}^{(f)}.
\]
If $F\ge2$ families survive, their representatives then compete on equal terms through
\[
h_f^{\rm cross}=\frac1F\sum_{q=1}^FH_{fq}^{\rm cross},
\]
with self-comparisons set to zero but kept in the denominators. The ARI and Jaccard components of the winning representative $f^\star$ are kept separately for later use,
\[
A_{r,g,\alpha}=\frac1F\sum_{q=1}^FA_{f^\star q},
\qquad
J_{r,g,\alpha}=\frac1F\sum_{q=1}^FJ_{f^\star q},
\]
and define the candidate score
\begin{equation}
H_{r,g,\alpha}=A_{r,g,\alpha}+0.2J_{r,g,\alpha}.
\label{eq:hier-score}
\end{equation}
Because the zero diagonal stays in the denominator, the score is bounded by $1.2(F-1)/F$ and therefore reflects how much cross-family evidence is available. If only one family has valid fits, its within-family score is used. Ties are handled as described in Appendix~\ref{sec:supp-choice}.

\paragraph{TRKM: robust-view cross-validation (RV--CV).}
For TRKM, pairwise agreement between starts is less informative, because the reconstruction criterion can produce competing partitions with different implications in the ambient space. Instead, every terminal fit is tested in the same two robust geometries, a SIGN--PCA view and a WINSOR--PCA view, of common dimension
\[
d=\max\{1,\min(10,p,n-1,2\max(r,g-1))\}.
\]
The views are built once per candidate $(X,r,g)$ and applied to every terminal fit regardless of its family. The question is whether a fit's labels can be recovered by its own centroids. Five deterministic folds are formed cyclically by observation index. In each fold and view, centroids are computed from the retained training observations using the candidate labels, and the retained test observations are reassigned to the nearest training centroid. Agreement for fit $j$ in view $v$ and fold $t$ is
\[
\operatorname{Agr}_{jvt}=
\frac{1}{|I^{\rm test}_{jt}|}
\sum_{i\in I^{\rm test}_{jt}}1\{\widehat z_{jvt}(i)=z_{ij}\},
\]
and the score is
\begin{equation}
\mathrm{RV}\text{--}\mathrm{CV}_j
=\frac12\sum_{v\in\{\mathrm{SIGN},\mathrm{WINSOR}\}}
\operatorname{mean}_{t\in\mathcal T_j}\operatorname{Agr}_{jvt},
\label{eq:rvcv}
\end{equation}
where $\mathcal T_j$ is the set of usable folds. The fit with the largest RV--CV represents the candidate. Rules for degenerate folds and views and for ties are given in Appendix~\ref{sec:supp-choice}. RV--CV is an internal reproducibility check on a single fit; it does not cross-validate the whole procedure and does not pool information across starts.

\subsection{Comparing structures along the trimming path}\label{sec:method-path}

After this step each $(r,g,\alpha)$ has one representative fit. The grids are
\begin{gather*}
\mathcal P=\{(r,g):1\le r<g\le6\},\\
\mathcal A=\{0,0.025,0.05,0.075,0.10,0.125,0.15,0.20,0.25,0.30\},
\end{gather*}
so each data set yields 150 representatives per method.

Two kinds of information are available at this point and should not be confused: how well a fit represents the retained data, and how strongly the solution is supported. For the first, let
\[
TSS=\|X_I-\bar X_I\|_F^2,\qquad
subTSS=\|(X_I-\bar X_I)U\|_F^2,
\qquad
W_{\rm ratio}=subTSS/TSS.
\]
For TFKM we record
\[
R_F^2=1-\frac{J_{\rm trim}}{subTSS},
\]
and for TRKM, with retained reconstruction residual $RSS_{\rm recon}$,
\[
R_R^2=1-\frac{RSS_{\rm recon}}{TSS},
\]
so that $R_R^2=W_{\rm ratio}R_F^2$ for a common state. HIER4 and RV--CV provide the second kind of information.

At each $\alpha$, TFKM selects the $(r,g)$ with the largest HIER4 score~\eqref{eq:hier-score}, breaking exact ties in favor of smaller $r+g$, then smaller $g$, then smaller $r$. This gives one sequence of per-level winners $w_\alpha$, used both to decide whether to trim and to start the simplification step.

TRKM uses two per-level winners, because its reconstruction and reproducibility scores answer different questions. The \emph{reproducibility winner} $w^{\rm RV}_\alpha$ maximizes RV--CV over the 15 pairs, with ties broken by larger $R_R^2$ and then by the parsimony order above; its path describes how reproducible the preferred structure is at each trimming level. The \emph{reconstruction-constrained winner} $w^{\rm R}_\alpha$ first restricts attention to structures whose reconstruction is within $0.01$ of the best,
\[
\mathcal N_{0.01}(\alpha)=
\left\{(r,g):R_R^2(r,g,\alpha)\ge
\max_{(r',g')\in\mathcal P}R_R^2(r',g',\alpha)-0.01\right\},
\]
and then takes the pair in this set with the largest RV--CV, with the same parsimony order for ties. The tolerance prevents a negligible gain in reconstruction from overriding much stronger reproducibility evidence. The first path is used to decide whether to trim (Section~\ref{sec:method-trim}), the second to start the simplification step (Section~\ref{sec:method-backward}).

For both methods, the \emph{modal} structures are those that occur most often along a path of winners. A structure $(r,g)$ is \emph{componentwise maximal} in a set when no other element $(r',g')$ of the set has $r'\ge r$ and $g'\ge g$ with at least one strict inequality.

\subsection{Deciding whether to trim}\label{sec:method-trim}

Any positive $\alpha$ on the grid can be fitted; that does not make the excluded observations a separate population. Positive trimming is adopted only when there is direct evidence of an outer population, or when the trimming path shows that excluding observations changes the fit in a coherent and reproducible way. This matters most for bridge contamination, where atypical observations lie inside the group geometry and do not form a removable set. Both procedures use the same two-step logic: a radial diagnostic computed from the data alone, and, if it does not fire, a method-specific rule based on the trimming path.

\paragraph{Radial evidence (both methods).}
Leverage that perturbs many coordinates at once can show up as a separate radial population even when the partition is uncertain. With $m_j$ and $s_j$ the coordinatewise median and (floored) MAD of the original variables, let
\[
Z_{ij}=\frac{X_{ij}-m_j}{s_j},
\qquad
q_i=\frac1p\sum_{j=1}^pZ_{ij}^2,
\qquad
\ell_i=\log\{\max(q_i,\varepsilon_s)\}.
\]
One- and two-component Gaussian mixtures are fitted to the log-radial scores $\ell_i$. If $\Delta BIC=BIC_1-BIC_2>0$ and the component with the higher mean has weight $0<w_H\le 0.30$, that weight is rounded to the nearest positive grid point to give $\widehat\alpha_{\rm rad}$. (The MAD floor and the minimum number of finite scores are given in Appendix~\ref{sec:supp-radial}.) Because this diagnostic depends only on $X$, it returns the same decision for TFKM and TRKM.

\paragraph{Path evidence for TFKM: unstable assignments, stable agreement on what to trim.}
The componentwise-maximal modal structures of the TFKM path $w_\alpha$ are examined one at a time. For such a mode $m$, let $A_m(\alpha)$ and $J_m(\alpha)$ be the cross-family ARI and Jaccard components in~\eqref{eq:hier-score} at level $\alpha$, and summarize the path by
\[
A_{{\rm med},m}=\operatorname{median}_{\alpha\in\mathcal A}A_m(\alpha),
\qquad
J_{\max,m}=\max_{\alpha>0}J_m(\alpha),
\]
\[
D_{m}=\frac{\max_\alpha A_m(\alpha)-\min_\alpha A_m(\alpha)}{A_m(0)},
\]
where $D_m$ is computed only when $A_m(0)>0$. Trimming is credible when the search families largely agree on \emph{which} observations to remove, while the assignments themselves are not uniformly stable along the path. Mode $m$ supports positive trimming when
\begin{equation}
J_{\max,m}\ge 0.50
\quad\text{and}\quad
\{A_{{\rm med},m}<0.44\ \text{or}\ D_{m}\ge 0.12\},
\label{eq:tfkm-action}
\end{equation}
and the proposed level $\widehat\alpha_{\rm path}$ is the smallest positive $\alpha$ maximizing $J_m(\alpha)$. If several modes are examined, they must agree on the decision and, when positive, on the level; otherwise positive trimming is not adopted and the output is flagged as ambiguous.

\paragraph{Path evidence for TRKM: reproducibility after trimming.}
For TRKM the rule asks whether the reproducibility of the modal structure improves materially only after some trimming. The modal structure is taken from the path of reproducibility winners $w^{\rm RV}_\alpha$; if several structures are equally frequent, the one with the largest $r+g$, then $g$, then $r$ is used. With $RV(\alpha)$ the RV--CV value of this structure at level $\alpha$, let
\[
\alpha_{0.01}=\min\Bigl\{\alpha:RV(\alpha)\ge\max_{\alpha'\in\mathcal A}RV(\alpha')-0.01\Bigr\},
\quad
\Delta RV=\max_\alpha RV(\alpha)-\min_\alpha RV(\alpha).
\]
The rule proposes $\widehat\alpha_{\rm path}=\alpha_{0.01}$ when $\alpha_{0.01}>0$ and $\Delta RV\ge 0.025$.

\paragraph{Composition.}
Because the radial branch is direct evidence of an outer population, it takes precedence in both procedures. With $\widehat\kappa\in\{0,1\}$ indicating whether positive trimming is adopted,
\begin{equation}
(\widehat\kappa,\widehat\alpha)=
\begin{cases}
(1,\widehat\alpha_{\rm rad}),&\text{if the radial condition holds},\\
(1,\widehat\alpha_{\rm path}),&\text{if the radial condition fails and the method's path rule holds},\\
(0,\,\cdot\,),&\text{otherwise}.
\end{cases}
\label{eq:trim-composition}
\end{equation}
The radial branch looks for a diffuse outer population, whereas the path rules ask whether the cluster geometry changes coherently once observations are excluded. Bridge observations need satisfy neither condition. When neither fires, the evidence does not support trimming; this is not a claim that the sample is uncontaminated.

\subsection{Simplifying without changing solution}\label{sec:method-backward}

Once positive trimming is adopted, the selected state may still be more complex than needed. We simplify it backwards, but only when the simpler state represents essentially the same grouping. The criterion is agreement on observations retained by both states rather than a small improvement in a local objective, so each backward move is a continuity check.

Both procedures start from the upper part of the trimming path. Once $\widehat\alpha$ is chosen, the per-level winners at $\alpha\ge\widehat\alpha$ are collected ($w_\alpha$ for TFKM, $w^{\rm R}_\alpha$ for TRKM), and their componentwise-maximal elements are the starting states. From each starting state the procedure follows the predecessor map
\[
b(r,g)=\begin{cases}
(r,g-1),&g>r+1,\\
(r-1,g-1),&g=r+1\text{ and }r>1,
\end{cases}
\]
down to $(1,2)$, refitting at $\widehat\alpha$. Every move lowers $g$ by one, and lowers $r$ as well when $r=g-1$. A move is accepted only if the fits at the two states have ARI at least $0.90$ on their mutually retained observations. The fit used to accept a move is carried forward and becomes the current fit for the next move, so that a path cannot look stable merely because it switches to a different local solution at each step. The first move that fails the $0.90$ requirement stops the path. If different starting states end at different structures, the output is flagged as structurally ambiguous.

The two methods differ only in how the fits at the two ends of a move are chosen from the corresponding start pools. TFKM ranks fits by their support across the two states and by cross-family HIER4 evidence, and uses the fitted objective only afterwards. TRKM restricts each pool to the initialization family of its RV--CV-selected fit and ranks pairs of fits by cross-state support and then by RV--CV. The exact ranking keys are given in Appendix~\ref{sec:supp-backward}.

\subsection{What the procedures return}\label{sec:method-output}

Both procedures report the selected structure $(r,g)$, the trimming decision with its level when positive, and an ambiguity flag. When positive trimming is adopted, the structure is the end point of the simplification step, and the fit reached there supplies a terminal partition and trimming mask. When no trimming is adopted, no simplification is run: TFKM reports its componentwise-maximal modal structure, flagged as ambiguous if there is more than one, and TRKM reports the modal structure of its reproducibility path. In this case no terminal fit is constructed, because no single trimming level has been selected.

For evaluation we therefore keep three targets apart:
\[
T_S=(r_0,g_0),\quad
T_A=A_\star\in\{0,1\},\quad
T_\alpha=\alpha_0\ \text{(when positive trimming is identifiable)}.
\]
A data set counts as jointly correct when $(r_0,g_0,A_\star)$ is recovered if $A_\star=0$, and $(r_0,g_0,A_\star,\alpha_0)$ if $A_\star=1$. In the simulations $A_\star=1$ for diffuse leverage and $A_\star=0$ for clean and bridge data, so the nominal bridge fraction is not a trimming target. Agreement with the generating labels and exact recovery of the contamination mask are reported as additional outcomes.

\section{Simulation study}\label{sec:simulation}

The simulation study asks two questions. Do the complete procedures recover the structure and the trimming decision across dimensions, noise levels and contamination types? And when they fail, does the failure come from the data, from the search or from the selection rule? Generating labels, contamination indicators, subspaces and structural parameters are never available to the procedures; they are used only for evaluation.

\subsection{Design}\label{sec:sim-design}

All data sets have $n=120$ observations in three panels.
\begin{itemize}[leftmargin=*]
\item \textbf{MAIN} fixes $(r_0,g_0)=(3,4)$ and mean pairwise centroid distance $\delta=10$, and crosses $p/n\in\{0.5,1,2\}$ with orthogonal noise variance $\rho\in\{0.25,1,2\}$ under clean, bridge and leverage contamination (27 cells).
\item \textbf{STRUCTURE} fixes $p/n=1$, $\rho=1$, $\delta=10$ and varies $(r_0,g_0)\in\{(2,3),(3,5),(4,5)\}$ under the three contamination types (9 cells). In the $(3,5)$ case $r_0<g_0-1$, so the rank is not determined by the number of groups.
\item \textbf{WEAK} keeps $(3,4)$ under leverage contamination but lowers separation to $\delta=7$, with $(p/n,\rho)\in\{(1,1),(1,2),(2,0.25),(2,1),(2,2)\}$ (5 cells). It is designed so that the statistical signal remains while the search becomes difficult.
\end{itemize}
Each of the 41 cells has 30 replications, giving 1230 data sets per method.

\begin{figure}[htbp]
\centering
\includegraphics[width=.92\linewidth]{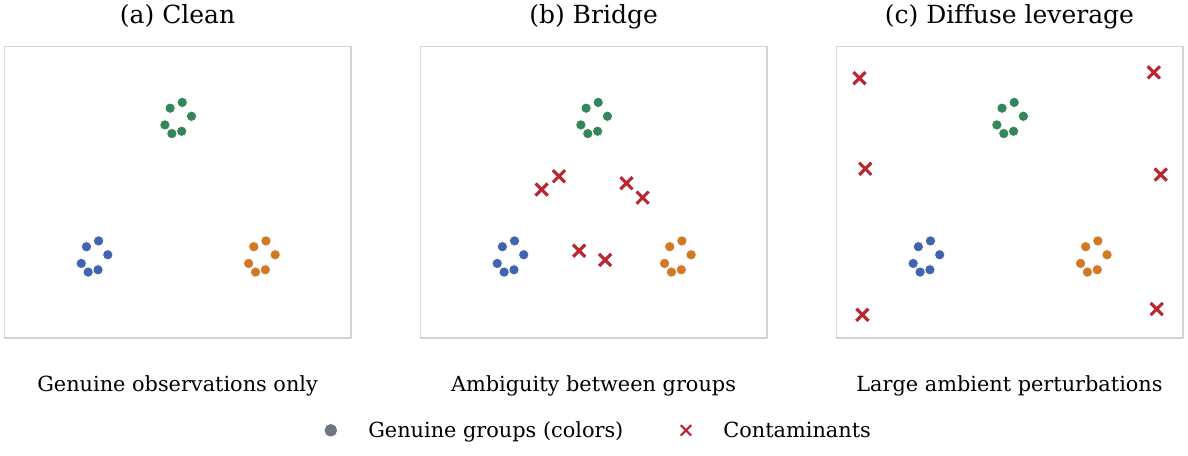}
\caption{Contamination geometries in the simulation study (schematic). Colors mark three genuine groups and crosses mark contaminants. Bridge points lie between groups; diffuse leverage perturbs many ambient coordinates.}
\label{fig:contamination-geometries}
\end{figure}

Genuine observations, that is, observations that are not contaminants, follow the latent model
\begin{equation}
X_i=U_0\bigl(\mu_{Y_i}+\varepsilon_i\bigr)+V_0\eta_i,\qquad
\varepsilon_i\sim N(0,I_{r_0}),\quad \eta_i\sim N(0,\rho I_{p-r_0}),
\label{eq:dgp}
\end{equation}
so that $X_i\mid Y_i=k\sim N\bigl(U_0\mu_k,\;U_0U_0'+\rho(I_p-U_0U_0')\bigr)$. Here $[U_0,V_0]$ is a random $p\times p$ orthonormal matrix obtained by QR decomposition, the labels $Y_i\in\{1,\ldots,g_0\}$ form balanced groups, and the centroids $\mu_1,\ldots,\mu_{g_0}\in\mathbb R^{r_0}$ are the rows of the $g_0\times r_0$ matrix of the first $r_0$ left singular vectors of $I_{g_0}-\mathbf 1\mathbf 1'/g_0$, rescaled to mean pairwise distance $\delta$. The noise variance is one in the signal coordinates and $\rho$ in the orthogonal complement. In contaminated samples, 12 of the 120 observations, chosen uniformly without replacement, become contaminants (Figure~\ref{fig:contamination-geometries}). A \emph{bridge} contaminant from group $k$ is replaced by
\[
\widetilde X_i=U_0\bigl(\tfrac12(\mu_k+\mu_{k+1})+2\zeta_i\bigr)+V_0\eta_i,\qquad \zeta_i\sim N(0,I_{r_0}),
\]
with $k+1$ taken cyclically, so it lies between two groups inside the signal space. A \emph{leverage} contaminant is perturbed in every coordinate,
\[
\widetilde X_{ij}=X_{ij}+10\,s_j\,\xi_{ij},\qquad \xi_{ij}\sim N(0,1),
\]
where $s_j$ is the sample standard deviation of variable $j$ before contamination; these points can pull the fitted subspace. Data matrices are column-centered; seeds are listed in Appendix~\ref{sec:supp-sim}.

The nominal contamination fraction is $0.10$ for bridge and leverage data. As explained in Section~\ref{sec:method-output}, positive trimming is the target only for leverage: bridge points need not be separable from the genuine groups by any trimming diagnostic, so clean and bridge data have a no-trimming target. Truth-assisted calculations at the \emph{generating state} use $(r_0,g_0)$, the generating labels and $\alpha_0$ equal to the nominal fraction ($0$ for clean data, $0.10$ otherwise) with the true contaminants as trimmed set.

To interpret success and failure we also compute \emph{truth-assisted} reference quantities from the generating labels and subspace, which the procedures never see. The most important is \emph{projected-oracle classification}: nearest-centroid assignment of the genuine observations in a subspace estimated from the same sample with the true labels. If it is exact, the projected geometry of that sample can support the correct grouping. Two summary margins are also used: $G_{\rm spec}$, the ratio of the boundary eigengap of an empirical reference operator to its sampling perturbation, and $G_{P,\mathrm{eff}}$, a conservative projected classification margin combining centroid separation, subspace overlap and orthogonal noise. Their exact definitions are in Appendix~\ref{sec:supp-reference}.

Because a no-trimming decision returns a structure without a terminal fit (Section~\ref{sec:method-output}), no overall ``selected ARI'' is reported. All ARI values below refer to explicitly defined fits or start pools.

\subsection{Recovery of structure and trimming decision}\label{sec:sim-overall}

Table~\ref{tab:primary-performance} reports, by contamination type, the fraction of data sets in which each procedure recovers the exact structure $(r_0,g_0)$, makes the correct trimming decision, and achieves both (for leverage data, also with the exact trimming level).

\begin{table}[htbp]
\centering
\caption{Automatic recovery by panel and contamination type. ``$(r,g)$'' is the fraction of data sets with exactly recovered structure; ``Trim'' the fraction of correct trim/no-trim decisions; ``Joint'' requires exact $(r,g)$, the correct decision and, for leverage, the exact trimming level $\alpha_0=0.10$. Results for each of the 41 design cells are in Table~\ref{tab:supp-cells}.}
\label{tab:primary-performance}
\small
\setlength{\tabcolsep}{4pt}
\begin{tabular}{llr ccc ccc}
\toprule
 & & & \multicolumn{3}{c}{REG-HD-TFKM} & \multicolumn{3}{c}{REG-HD-TRKM}\\
\cmidrule(lr){4-6}\cmidrule(lr){7-9}
Panel & Contamination & $N$ & $(r,g)$ & Trim & Joint & $(r,g)$ & Trim & Joint\\
\midrule
MAIN & Clean & 270 & 0.863 & 0.993 & 0.859 & 0.926 & 0.996 & 0.926\\
 & Bridge & 270 & 0.874 & 0.989 & 0.863 & 0.922 & 0.956 & 0.881\\
 & Leverage & 270 & 0.793 & 1.000 & 0.793 & 0.685 & 1.000 & 0.685\\
\addlinespace
STRUCTURE & Clean & 90 & 0.678 & 1.000 & 0.678 & 0.611 & 0.956 & 0.611\\
 & Bridge & 90 & 0.733 & 0.989 & 0.733 & 0.656 & 0.933 & 0.611\\
 & Leverage & 90 & 0.633 & 1.000 & 0.633 & 0.322 & 1.000 & 0.322\\
\addlinespace
WEAK & Leverage & 150 & 0.420 & 1.000 & 0.420 & 0.360 & 1.000 & 0.360\\
\midrule
All & Clean & 360 & 0.817 & 0.994 & 0.814 & 0.847 & 0.986 & 0.847\\
 & Bridge & 360 & 0.839 & 0.989 & 0.831 & 0.856 & 0.950 & 0.814\\
 & Leverage & 510 & 0.655 & 1.000 & 0.655 & 0.525 & 1.000 & 0.525\\
\addlinespace
\multicolumn{2}{l}{All data sets} & 1230 & 0.756 & 0.995 & 0.753 & 0.716 & 0.981 & 0.704\\
\bottomrule
\end{tabular}
\end{table}

Clean and bridge data behave similarly. Over all clean and bridge data sets, both procedures make the correct no-trimming decision in at least 95\% of data sets and recover the full structure in more than 80\% (0.86--0.93 in MAIN, 0.61--0.73 in STRUCTURE), so bridge points are not trimmed merely because a positive $\alpha$ is available on the grid. TFKM trims in 6 of these 720 data sets (2 through the radial branch, 4 through its path rule) and TRKM in 23, mostly through its path rule.

Under diffuse leverage the difficulty lies in the structure, not in the trimming decision. The radial diagnostic fires in all 510 leverage data sets and returns the exact level $\alpha_0=0.10$ every time, so both procedures trim correctly at the correct level. TFKM recovers the exact $(r_0,g_0)$ in 65.5\% of leverage data sets and TRKM in 52.5\%. The advantage of TFKM holds in every panel (MAIN 0.793 against 0.685, STRUCTURE 0.633 against 0.322, WEAK 0.420 against 0.360), and TFKM flags 10 of the 510 outputs as ambiguous against 55 for TRKM. The errors of TFKM go in both directions: the selected rank is too large in 13.5\% and too small in 18.4\% of leverage data sets, and the number of groups is too large in 11.0\% and too small in 16.3\%. One design is hard for both methods (see also the cell-level results in Table~\ref{tab:supp-cells}). When $(r_0,g_0)=(3,5)$, five centroids cannot be equidistant in three dimensions: the two closest groups are about 6.9 apart in the median data set, against 9.6--9.8 in the other designs. Without contamination, TFKM recovers $(3,5)$ in 11 of 30 data sets and TRKM in none; both select $(3,4)$, merging the two closest groups, in 19 of 30. Under leverage the number of groups is usually right but the rank is not. TFKM keeps $g=5$ in 20 of 30 data sets, yet recovers $(3,5)$ in only 2 and selects $(4,5)$ in 18; TRKM recovers $(3,5)$ in 6. The backward path of TFKM usually starts at $(5,6)$, accepts the move to $(4,5)$ and stops there, because the next state $(3,4)$ merges two groups. It never visits $(3,5)$, since every move lowers $g$ (Section~\ref{sec:method-backward}). Moves that lower the rank at fixed $g$ would need their own acceptance criterion: a partition that is preserved in fewer dimensions does not show that the dropped direction carries no group signal, so the ARI criterion would tend to accept rank reductions also when the rank is correct.

Good partitions are usually available: at the generating parameters the representative fits have mean ARI on genuine observations of 0.950 (TFKM) and 0.930 (TRKM), and the best representative over all 15 $(r,g)$ pairs at $\alpha_0$, whatever its structure, reaches 0.984 and 0.977. The harder task is selecting the exact structure, and the rest of this section examines where it fails.

\subsection{Where accuracy is lost: geometry, search and selection}\label{sec:sim-weak}

\begin{figure}[htbp]
\centering
\includegraphics[width=.94\linewidth]{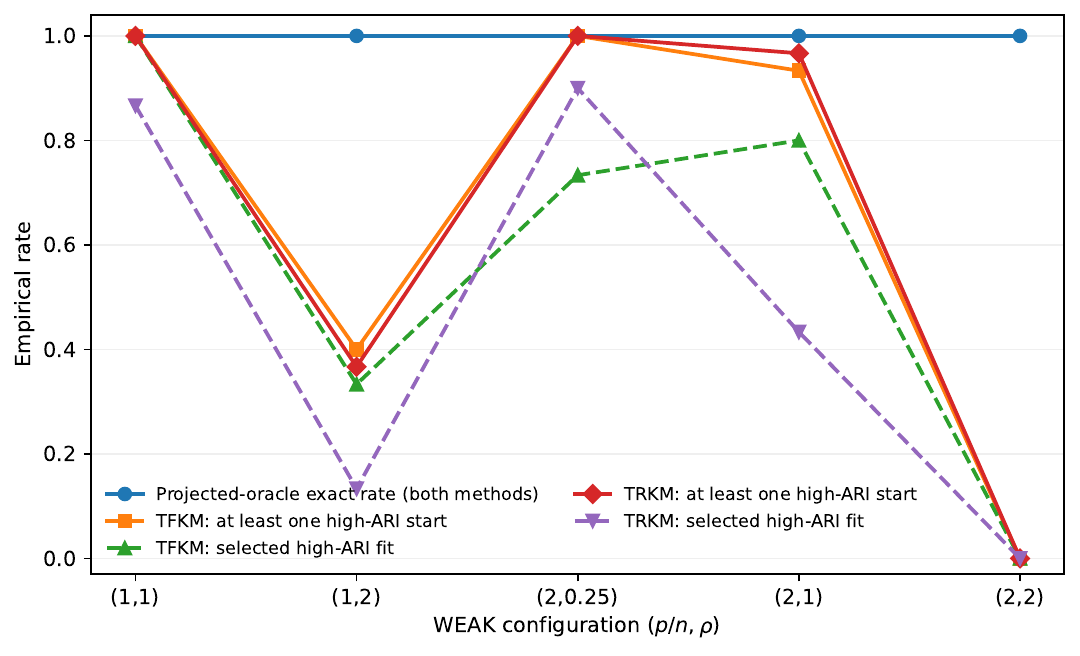}
\caption{WEAK panel: projected geometry versus multistart access and selection. ``At least one high-ARI start'' is the fraction of data sets whose 100-start pool contains a fit with genuine-observation ARI $\ge 0.90$; ``selected high-ARI fit'' is the unconditional fraction in which the choice rule returns such a fit. Projected-oracle classification is exact throughout.}
\label{fig:geometry-access}
\end{figure}

The WEAK panel separates three conditions that are usually merged into one:
\begin{enumerate}[leftmargin=*]
\item \emph{geometry}: the projected data can support the correct grouping (projected-oracle classification is exact);
\item \emph{access}: at least one of the 100 starts at the generating $(r_0,g_0,\alpha_0)$ ends in a high-quality fit, defined as ARI at least $0.90$ on genuine observations;
\item \emph{selection}: the method's own choice rule (HIER4 for TFKM, RV--CV for TRKM) returns such a fit.
\end{enumerate}
Table~\ref{tab:weak-access} and Figure~\ref{fig:geometry-access} report these conditions for each WEAK cell.

\begin{table}[htbp]
\centering
\caption{WEAK panel: projected geometry, access to a good solution and selection. ``Available'' is the fraction of data sets whose 100-start pool contains a fit with genuine-observation ARI $\ge 0.90$; ``Selected'' is the unconditional fraction in which the choice rule returns such a fit. Projected-oracle classification is exact in every row.}
\label{tab:weak-access}
\small
\begin{tabular}{llrrrr}
\toprule
Method & $(p/n,\rho)$ & Median $G_{\rm spec}$ & Median $G_{P,\mathrm{eff}}$ & Available & Selected\\
\midrule
TFKM & $(1,1)$   & 0.851 & 0.053 & 1.000 & 1.000\\
TFKM & $(1,2)$   & 0.851 & 0.056 & 0.400 & 0.333\\
TFKM & $(2,0.25)$ & 0.849 & 0.184 & 1.000 & 0.733\\
TFKM & $(2,1)$   & 0.850 & 0.417 & 0.933 & 0.800\\
TFKM & $(2,2)$   & 0.851 & 0.284 & 0 & 0\\
\addlinespace
TRKM & $(1,1)$   & 1.556 & 0.067 & 1.000 & 0.867\\
TRKM & $(1,2)$   & 1.444 & 0.073 & 0.367 & 0.133\\
TRKM & $(2,0.25)$ & 1.623 & 0.186 & 1.000 & 0.900\\
TRKM & $(2,1)$   & 1.365 & 0.418 & 0.967 & 0.433\\
TRKM & $(2,2)$   & 1.204 & 0.285 & 0 & 0\\
\bottomrule
\end{tabular}
\end{table}

Geometry is never the limiting condition here: projected-oracle classification is exact in every WEAK data set, for both methods. Each of the other two conditions is limiting in some cell.
\begin{itemize}[leftmargin=*]
\item At $(p/n,\rho)=(2,2)$ no start reaches a good solution for either method. The data support the grouping, but the search does not find it.
\item At $(2,0.25)$ a good solution is always in the pool, yet TFKM selects it in 73\% and TRKM in 90\% of data sets. The loss is due to selection alone.
\item At $(1,2)$ both mechanisms contribute: a good solution is available in only 37--40\% of data sets, and is selected in even fewer (33\% for TFKM, 13\% for TRKM).
\item At $(2,1)$ TRKM almost always reaches a good solution (0.967) but selects it in only 43\% of data sets.
\end{itemize}
The summary margins do not predict these outcomes reliably. The conservative condition $G_{P,\mathrm{eff}}>1$ is never met in WEAK, even though projected-oracle classification is always exact; over the whole study, projected-oracle classification is exact in 1230/1230 TRKM and 1228/1230 TFKM data sets. Within the WEAK panel, $G_{\rm spec}$ predicts the availability of a good TRKM solution well (AUC $0.900$) but not of a TFKM solution (AUC $0.601$). Favorable geometry is therefore not sufficient: it says little about whether a finite search will reach and recognize the right solution.

\subsection{A good partition is not a recovered state}\label{sec:sim-local}

The previous results distinguish high ARI from exact recovery of the structure. The same distinction appears at the level of a single update. At the generating state, assigning each genuine observation to the nearest sample centroid of the true groups in the estimated subspace is exact in all 1230 data sets for both methods. (Projected-oracle classification, which uses reference centroids instead, misses two TFKM data sets.) One update returns the genuine labels and the complete trimming mask in 1180 TFKM and 1181 TRKM data sets (about 96\%). The shortfall is concentrated in MAIN bridge cells (rates 0.819 and 0.822). There, the smallest residual of a contaminant exceeds the largest residual of a genuine observation in the median data set, but not in the lowest decile, so the trimming step occasionally swaps a contaminant for a genuine observation.

\begin{figure}[htbp]
\centering
\includegraphics[width=.82\linewidth]{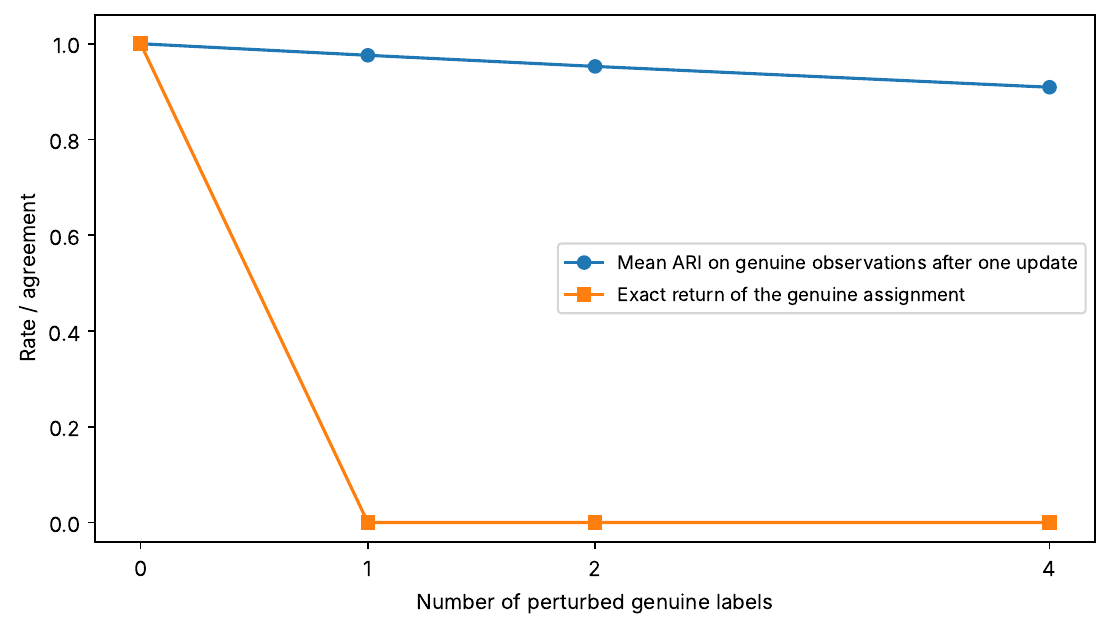}
\caption{Exact recovery and partition agreement are different targets. Generating labels of genuine observations near group boundaries are perturbed at the generating state before one complete update; results are pooled over both methods and all 1230 data sets per method. Exact assignment return is zero for every positive perturbation, while mean ARI on genuine observations remains high.}
\label{fig:partition-vs-exact}
\end{figure}

To see how quickly exact recovery is lost, we perturbed the generating state before one complete update, changing 1, 2 or 4 genuine labels selected near a group boundary (Figure~\ref{fig:partition-vs-exact}). After any positive number of changed labels, the update never returns the exact assignment, yet the mean ARI on genuine observations remains $0.976$, $0.953$ and $0.909$. Perturbing only trimming decisions gives the same picture: the exact mask is never returned, while mean ARI stays between $0.995$ and $0.996$. A practically excellent partition therefore need not correspond to an exactly recovered state, and an evaluation based on ARI alone cannot reveal whether the structure has been identified.

\subsection{Bridge and leverage contamination}\label{sec:sim-trim}

The two contamination types differ in a way that justifies treating the trimming decision separately from the nominal contamination fraction. Bridge points lie between centroids inside the signal space, so in the generating subspace $U_0$ their residuals are smaller than those of many genuine observations: in MAIN bridge cells the median gap between the smallest contaminant residual and the largest genuine residual lies between about $-7.2$ and $-5.7$, and ordering by these residuals recovers the bridge set exactly in only 7\% of data sets. In the subspace estimated at the generating state, from which the contaminants are excluded, they usually have the largest residuals (Section~\ref{sec:sim-local}); the procedures cannot rely on this, because they do not know which observations to exclude. The genuine partition, in contrast, is recovered exactly at the generating state. Bridge contamination can therefore leave the grouping fully recoverable while the contaminant set is not identifiable from residuals.

Diffuse leverage is separable. As noted above, the radial diagnostic detects it in all 510 leverage data sets, with 2 positives among the 720 clean or bridge data sets; both the BIC contrast and the standardized distance between the two fitted component means separate leverage from other data sets perfectly (AUC one). These are detection rates at the level of data sets, not of individual observations, and the design does not locate a detection boundary.

\subsection{Known-parameter benchmark}\label{sec:sim-benchmark}

To check whether the fitting cores themselves limit accuracy, we compared seven estimators on the same 1230 data sets when the true parameters are supplied: $(r_0,g_0,\alpha_0)$ for the subspace methods and $(g_0,\alpha_0)$ for trimmed K-means (TKM) and TCLUST, which have no rank parameter. For REG-HD-TFKM and REG-HD-TRKM these fits are the candidates at $(r_0,g_0,\alpha_0)$ already computed for the selection, so the comparison concerns exactly the fits that the selection rules compare. All methods use 100 starts; REG-HD-TFKM chooses among them with HIER4 and REG-HD-TRKM with RV--CV. TKM and TCLUST use the FSDA implementations \citep{riani2012fsda}, TCLUST with the spherical (VII) restriction. The three tandem methods apply TKM to the first $r_0$ principal components: classical (PCA--TKM), spatial-sign \citep[SSPCA--TKM;][]{locantore1999robust} and ROBPCA \citep[ROBPCA--TKM;][]{hubert2005robpca}, the last computed with the LIBRA implementation \citep{verboven2005libra} at its default coverage $0.75$. ARI is computed on all genuine observations, with those trimmed by a fit forming a separate class; for the regularized fits it is therefore slightly lower than the values in Section~\ref{sec:sim-overall}, which keep the nearest-group labels of trimmed genuine observations.

\begin{table}[htbp]
\centering
\caption{Known-parameter benchmark on the 1230 data sets of the simulation study: mean ARI on genuine observations by contamination type ($N=360$, 360 and 510), over all data sets, and mean runtime per fit.}
\label{tab:competitors}
\small
\begin{tabular}{lrrrrr}
\toprule
Method & Clean & Bridge & Leverage & All & Runtime (s)\\
\midrule
TKM & 1.000 & 0.902 & 0.933 & 0.943 & 0.16\\
PCA--TKM & 1.000 & 0.966 & 0.246 & 0.678 & 0.09\\
SSPCA--TKM & 1.000 & 0.967 & 0.967 & 0.976 & 0.09\\
ROBPCA--TKM & 1.000 & 0.966 & 0.962 & 0.974 & 0.56\\
TCLUST & 1.000 & 0.902 & 0.923 & 0.939 & 8.25\\
REG-HD-TFKM & 1.000 & 0.961 & 0.883 & 0.940 & 6.25\\
REG-HD-TRKM & 0.996 & 0.952 & 0.838 & 0.918 & 5.60\\
\bottomrule
\end{tabular}
\end{table}

On clean data every estimator recovers the partition (mean ARI at least 0.996). Bridge contamination favors the subspace methods ($0.952$--$0.967$ against $0.902$ for TKM and TCLUST): trimming in the full $p$-dimensional space, TKM and TCLUST mostly remove genuine observations rather than bridge points (mean trimming $F_1$ of 0.39, against 0.70--0.78 for the subspace methods). Diffuse leverage separates the methods more sharply. Classical PCA--TKM fails (mean ARI 0.246) because the contaminants determine the leading principal components, while TKM, TCLUST and the two robust tandem methods reach $0.923$--$0.967$.

The regularized cores are less accurate under leverage (0.883 for REG-HD-TFKM and 0.838 for REG-HD-TRKM). The shortfall is concentrated where the orthogonal noise is large: in the MAIN leverage cells with $\rho=2$ their mean ARI is 0.894 and 0.789, against at least 0.994 for TKM, TCLUST and the robust tandem methods; in the WEAK panel all methods lose accuracy (0.741--0.895 for these four competitors, 0.751 and 0.616 for the regularized cores). The geometry is not the cause. In each of the 115 TFKM and 124 TRKM leverage data sets in which the regularized fit at the true parameters has ARI below 0.90, one update from the generating state returns its genuine labels and trimming mask exactly (Section~\ref{sec:sim-local}). The generating state is therefore a fixed point of the algorithm that the 100-start search did not reach, or reached but did not choose: these are the access and selection losses of Section~\ref{sec:sim-weak}, here with the parameters fixed. These data sets account for much of the structural error under leverage in Table~\ref{tab:primary-performance}: the exact $(r_0,g_0)$ is selected in 19\% of them for TFKM and 29\% for TRKM, against 79\% and 60\% in the other leverage data sets. When the fit at the true parameters is poor, the selection rules have little reason to prefer it, so the start pool and the choice among starts are the components to strengthen.

A regularized fit with 100 starts takes 5.9 seconds on average at $n=120$, against at most 0.56 seconds for TKM and the tandem methods and 8.2 seconds for TCLUST. Selecting $(r,g,\alpha)$ requires the 150 candidate fits of Section~\ref{sec:method-path}, which are independent and can run in parallel, plus the start pools regenerated by the simplification step.

\subsection{Is the group-number and rank signal present?}\label{sec:sim-signal}

A structural error can also arise because the sample does not contain a clear signal for $g$ or $r$. Two truth-assisted diagnostics, detailed in Appendix~\ref{sec:supp-sim}, indicate that the signal is present in this design. A distortion path computed on the genuine observations in the generating subspace has a clear change point at $g_0$ in all 1230 data sets: every gain before $g_0$ exceeds the first gain after $g_0$, with median ratios between 19 and 52 across panel and contamination groups. A cross-fitted between-group scatter recovers $r_0$ in all 1230 data sets by a midpoint eigenvalue rule. Both diagnostics use the true labels and are not selectors. They suggest that the structural errors of Section~\ref{sec:sim-overall} come from learning $(r,g)$ through a non-convex fit rather than from an absence of signal.

The design also shows when errors in $g$ become more frequent. They grow as the closest groups become less separated relative to the unit within-group spread: $g$ is correct in fewer than half of the WEAK data sets, where $\delta=7$ (0.453 for TFKM and 0.407 for TRKM), and in the $(3,5)$ design, whose closest groups are about 6.9 apart, both methods merge two groups in 19 of 30 clean data sets (Section~\ref{sec:sim-overall}). Under leverage they also grow with the orthogonal noise: in the MAIN panel, TFKM selects the correct $g$ in 0.989 of the data sets with $\rho=0.25$ and in 0.622 with $\rho=2$. Within a fixed design, however, they are not predicted by the geometry of the sample. In the MAIN panel, where $(r_0,g_0)=(3,4)$ and $\delta=10$ throughout, $g$ is wrong in 14.4\% (TFKM) and 13.7\% (TRKM) of data sets, yet neither the distance between the two closest sample group means nor the change-point ratio separates these data sets from the others (AUC between 0.48 and 0.57, where 0.50 is chance). Beyond the effects of separation and noise built into the design, errors in $g$ therefore arise in learning the structure through the non-convex fit and the selection rule.

\subsection{Summary}\label{sec:sim-summary}

The simulations support four conclusions. First, when the parameters are known both cores match established robust methods on clean and bridge data, but under diffuse leverage with large orthogonal noise their 100-start search sometimes misses the generating state, although that state is a fixed point. When the parameters are not known, the trimming decision is correct in 99.5\% (TFKM) and 98.1\% (TRKM) of data sets, and structure and trimming are recovered jointly in 81--85\% of clean and bridge data sets. Second, diffuse leverage is always detected and trimmed at the right level, but the structure is recovered in 65.5\% (TFKM) and 52.5\% (TRKM) of these data sets, with the weakest results in the weak-signal panel and when $r_0<g_0-1$. Third, favorable geometry, access to a good solution and its selection are distinct conditions, and in the WEAK panel each of the last two is limiting in some cell. Fourth, exact recovery of a state is a much stricter target than a high ARI.

\section{Application: euro-area bank balance sheets}\label{sec:application}

\subsection{Data and protocol}

We use confidential year-end 2014 supervisory data on euro-area banks under two representations of the same balance sheets. The first, ratios to total assets, follows the business-model study of \citet{farne2021banks}. The source matrix has 365 banks and 383 columns: log total assets and 382 granular items divided by total assets. Log total assets is held out from clustering and used only for interpretation. Two exact duplicate pairs are reduced to single observations, leaving a $363\times382$ matrix. The second representation contains 361 granular FINREP indicators on their original logarithmic scale; after removal of one repeated institution the matrix is $363\times361$. No further standardization is applied, and exact zeros, about 79\% of the entries, are kept. Ratios describe the composition of the balance sheet; log levels keep absolute scale and the breadth of reported positions.

Both representations are analyzed with the grids and rules used in the simulation study, without application-specific tuning: $1\le r<g\le6$ and $\alpha\in\{0,0.025,0.05,0.075,0.10,0.125,0.15,0.20,0.25,0.30\}$. Bank identifiers and the business-model labels of \citet{farne2021banks} play no role in fitting or selection. When a procedure finds no evidence for positive trimming it returns a structure but no terminal fit (Section~\ref{sec:method-output}); to aid interpretation we then show the $\alpha=0$ fit at the selected structure and call it \emph{descriptive}. Table~\ref{tab:bank-summary} summarizes the four selections.

\begin{table}[htbp]
	\centering
	\caption{Banking application: selected structures and trimming decisions. Descriptive partitions are $\alpha=0$ fits at the selected structure, shown when the procedure does not support positive trimming.}
	\label{tab:bank-summary}
	\small
	\begin{tabular}{llcccl}
		\toprule
		Representation & Method & Trimming & $\widehat\alpha$ & $(\widehat r,\widehat g)$ & Group sizes (trimmed) \\
		\midrule
		Ratios to assets & TFKM & none     & --   & $(2,3)$ & 328/8/27 (0), descriptive \\
		& TRKM & positive & 0.20  & $(1,2)$ & 286/5 (72) \\
		\addlinespace
		Log-level FINREP & TFKM & positive & 0.25  & $(1,2)$ & 108/165 (90) \\
		& TRKM & none     & --   & $(1,2)$ & 70/293 (0), descriptive \\
		\bottomrule
	\end{tabular}
\end{table}

\subsection{Ratios to assets: similar partitions, different subspaces}

TFKM selects $(r,g)=(2,3)$ and finds no evidence for trimming; its descriptive fit has groups of 328, 8 and 27 banks. TRKM adopts $\widehat\alpha=0.20$ through its path rule (the radial diagnostic does not fire); simplification starts from $(2,3)$, and one accepted backward move leads to $(1,2)$. The TRKM fit retains 291 banks in groups of 286 and 5, and trims 72.

On the 291 banks retained by TRKM the two partitions agree closely, with ARI $0.827$. Of the 286 banks in the large TRKM group, 284 belong to the dominant TFKM group, and all five banks in the small TRKM group belong to the eight-bank TFKM group. Table~\ref{tab:bank-ratio} shows where the TRKM trimming falls: it removes 25 of the 27 banks in the third TFKM group, 3 of 8 in the second and 44 of 328 in the first.

\begin{table}[htbp]
	\centering
	\caption{Ratios to assets: TRKM trimming within the descriptive TFKM groups.}
	\label{tab:bank-ratio}
	\small
	\begin{tabular}{lrrr}
		\toprule
		TFKM descriptive group & Size & Trimmed by TRKM & Retained by TRKM \\
		\midrule
		1 & 328 & 44 & 284 \\
		2 & 8   & 3  & 5   \\
		3 & 27  & 25 & 2   \\
		\midrule
		Total & 363 & 72 & 291 \\
		\bottomrule
	\end{tabular}
\end{table}

The agreement in labels does not extend to the subspaces. The TRKM direction is almost orthogonal to the TFKM plane: their squared overlap is $2.56\times10^{-4}$, a principal angle of $89.1^\circ$. Both subspaces lie in the row spaces of the respective centered retained data, with negligible mass in the null space, so the discrepancy is not an artefact of arbitrary directions in the ambient null space. The same grouping is supported by two very different low-dimensional explanations.

Size explains little of either geometry. The correlations of log total assets with the two TFKM scores are $-0.229$ and $-0.119$, and with the TRKM score $0.225$, although the 72 banks trimmed by TRKM are larger on average (mean log total assets 21.01 against 19.37 for retained banks). The strongest contrasts are composition variables related to market activity, derivatives and hedging, market funding, and repo and deposit items. In both methods the leading variable is the F~10.00 item for interest-rate derivatives held as economic hedges, measured by total trading notional, followed by the carrying amount of financial liabilities held for trading. We read the ratio geometry as a broad balance-sheet core together with small configurations with higher market, derivative and funding intensity. We do not map these groups onto the business-model labels of \citet{farne2021banks}, which are neither inputs nor targets here.

\subsection{Log levels: same structure, different retained samples}

The log-level representation shows the opposite pattern. Both procedures select $(r,g)=(1,2)$, but they apply it to different banks. The radial diagnostic separates two components clearly, but the higher-mean component carries weight $0.992$, far above the admissible bound $w_H\le 0.30$, so it does not fire: the separation describes a dominant and a residual population rather than a removable outer component. TFKM adopts $\widehat\alpha=0.25$ through its path rule and retains 273 banks in groups of 108 and 165. This decision rests on $J_{\max}=0.512$, just above the cutoff $0.50$, and would not survive a cutoff of $0.55$ (Appendix~\ref{sec:supp-thresholds}), so it is the least robust selection of the application; the positive TRKM decision under ratios ($\Delta RV=0.059$ against a cutoff of $0.025$) is not sensitive in this way. TRKM finds no evidence for positive trimming; its descriptive fit has groups of 70 and 293.

The two one-dimensional clustering spaces are again nearly orthogonal (principal angle $86.2^\circ$), and on the 273 banks retained by TFKM the ARI between the two partitions is $-0.0025$. Table~\ref{tab:bank-level} explains this value: TFKM trims 69 of the 70 banks in the smaller TRKM group and 21 of the 293 in the larger one. On the TFKM-retained banks the TRKM partition therefore has margins 1 and 272, against TFKM margins of 108 and 165. With these margins the ARI can take only two values, $-0.0025$ or $0.0039$, depending on which TFKM group contains the single remaining member of the small TRKM group. The near-zero ARI is forced by trimming and says little about how the two partitions compare on the full sample.

\begin{table}[htbp]
	\centering
	\caption{Log-level FINREP: TFKM trimming within the descriptive TRKM groups.}
	\label{tab:bank-level}
	\small
	\begin{tabular}{lrrr}
		\toprule
		TRKM descriptive group & Size & Trimmed by TFKM & Retained by TFKM \\
		\midrule
		1 & 70  & 69 & 1   \\
		2 & 293 & 21 & 272 \\
		\midrule
		Total & 363 & 90 & 273 \\
		\bottomrule
	\end{tabular}
\end{table}

The descriptive TRKM axis is largely a size axis. Mean log total assets is 25.81 in the 70-bank group and 22.37 in the 293-bank group, and the absolute correlation between the TRKM score and log total assets is $0.645$. TFKM trims almost the whole 70-bank group before forming its partition. Within the banks it retains, mean log total assets is 23.22 in the 108-bank group and 21.67 in the 165-bank group, and the absolute score--size correlation is $0.450$, so the retained split is not a size split alone. Its largest contrasts combine differences in whether, and how much, banks report derivative, debt-security, repo and other market-based positions.

\subsection{What the two representations show together}

Table~\ref{tab:bank-cross} puts the two cases side by side. With ratios, the partitions agree while the selected structures and subspaces differ sharply. With log levels, the selected structures coincide while the subspaces are nearly orthogonal and the partitions are evaluated on different retained populations. Partition, subspace, structure and trimmed set are therefore four distinct descriptions of a fit, and reporting one ARI or one selected $(r,g)$ would hide most of what the procedures found. The contrast between representations is also substantive: ratios remove much of the scale effect and expose composition, whereas log levels keep size and reporting breadth. The same institutions can thus show different cluster geometry under different economic representations, which is the situation in which joint estimation of subspace and partition is most informative.

\begin{table}[htbp]
	\centering
	\caption{Cross-method comparison under the two representations. ARI is computed on the observations retained by the trimmed fit.}
	\label{tab:bank-cross}
	\small
	\begin{tabular}{lcccl}
		\toprule
		Representation & Trimmed fit & ARI & Angle & Main difference \\
		\midrule
		Ratios to assets & TRKM & $0.827$ & $89.1^\circ$ & structure and subspace \\
		Log-level FINREP & TFKM & $-0.0025$ & $86.2^\circ$ & retained population \\
		\bottomrule
	\end{tabular}
\end{table}

\section{Discussion}\label{sec:conclusion}

Robust subspace clustering in high dimension faces several difficulties at once: the clustering space, the number of groups and the retained fraction are unknown, the scatter matrices can be singular, and the criterion is non-convex. REG-HD-TFKM and REG-HD-TRKM address these difficulties by taking one decision at a time and recording the evidence behind each.

The simulations show why the decisions should be kept apart. Correct projected classification, access to a good fit, selection of that fit and recovery of $(r,g)$ are distinct requirements, and in the weak-signal experiment the search and the selection rule are each the limiting step in some regime. A good partition, moreover, does not imply that the rank, the number of groups or the trimmed set have been recovered.

The contamination experiments add a second lesson. Positive trimming should not be read as an estimate of a nominal contamination fraction. Bridge observations lie inside the group geometry and need not be separable by radial or residual ordering, so declining to trim them is the intended behavior. Diffuse leverage forms a separable outer population, and the radial diagnostic detects it in every simulated data set. A decision not to trim therefore states that the evidence does not support trimming; it does not certify that the sample is clean.

The same experiments show where the procedures remain weak. Once leverage is trimmed at the right level, the structure is recovered in 65.5\% (TFKM) and 52.5\% (TRKM) of these data sets, and in fewer than half in the weak-signal panel, where the search itself sometimes fails to reach a good solution. Part of the loss is structural. Every backward move lowers the number of groups, so a rank overestimated at the correct $g$ cannot be corrected: when the truth is $(3,5)$ and the selected state is $(4,5)$, the path moves to $(3,4)$, and TFKM recovers $(3,5)$ in only 2 of 30 leverage data sets. Simplification paths that also lower the rank at fixed $g$ would remove this restriction, but they need an acceptance criterion specific to the rank, because agreement between partitions cannot show that a dropped direction carries no group signal (Section~\ref{sec:sim-overall}). The known-parameter benchmark identifies a further source. Under diffuse leverage with large orthogonal noise, the fit at the true $(r,g,\alpha)$ is itself often poor, although one update from the generating state returns it exactly, and in these data sets the structure is rarely recovered. Where the fitting cores are accurate at the true parameters, most of the remaining loss arises in structural selection.

Further limitations should be noted. The thresholds of the selection, trimming and simplification rules are fixed (Section~\ref{sec:method}) and have not been calibrated across a wider range of data-generating mechanisms. For the trimming rules, the stored statistics of the simulation study show that wide intervals of thresholds give the same or nearly the same decisions (Appendix~\ref{sec:supp-thresholds}); the thresholds that act on the structure could not be varied without refitting. The simulations use $n=120$ and a single family of Gaussian designs, so an evaluation on independently specified designs and larger samples would strengthen the evidence. The 100-start budget is finite, and the weak-signal experiment shows that it can miss a good solution. Selecting $(r,g,\alpha)$ requires 150 candidate fits per data set, each from 100 starts; in the simulation study this took a median of 12 minutes per data set and method (7, 12 and 29 minutes for $p=60$, 120 and 240) and less than 42 minutes in 99\% of cases, although the candidates can be computed in parallel. The fixed-parameter benchmark is meant as a check on fitting accuracy rather than as a ranking of methods, and does not include competitors that select $g$ and $\alpha$ automatically \citep{garciaescudero2026choice}. Finally, the supervisory banking data are confidential, so public replication is limited to code and non-identifying aggregate results.

Several extensions follow. Search budgets could adapt to the diversity of the solutions found, the start pool could include the solutions of fast robust tandem fits, which are accurate under leverage in the benchmark, or starts whose trimmed set is the outer component identified by the radial diagnostic, larger samples would allow external validation at the selection stage, and uncertainty statements for $(r,g,\alpha)$ would distinguish a stable structural conclusion from a merely best-scoring one. On the theoretical side, the simulations show that the available sufficient conditions for exact projected classification and for rank recovery are far from tight (Section~\ref{sec:sim-weak} and Appendix~\ref{sec:supp-rank}), so sharper conditions would help to explain when the search and the selection succeed. The distinction between statistical geometry, access to a good solution and selection among competing solutions is not specific to FKM and RKM: it arises whenever structural selection is built on top of a non-convex robust clustering criterion.

\backmatter

\section*{Declarations}

\begin{itemize}
	\item \textbf{Funding.} Matteo Farn\`e is partly supported by the PRIN Project No.~20224CRB9E, CUP C53C24000730006, entitled \emph{Latent variable models and dimensionality reduction methods for complex data} (PI Prof.\ Paolo Giordani) and funded by the Italian Ministry of University and Research.
	\item \textbf{Acknowledgements.} We gratefully thank the Supervisory Statistics Division of the European Central Bank for allowing the use of confidential balance sheet data for research purposes.
	\item \textbf{Competing interests.} The authors have no competing interests to declare that are relevant to the content of this article.
	\item \textbf{Data availability.} The simulated data can be regenerated with the code below. The euro-area supervisory banking data are confidential and cannot be shared; the non-identifying aggregate results used in Section~\ref{sec:application} are included in the replication package.
	\item \textbf{Code availability.} MATLAB code for both procedures and a Python/MATLAB package that reproduces all tables and figures from archived results are available at: \url{https://anonymous.4open.science/r/reghd-tfkm-trkm}.
\end{itemize}

\begin{appendices}

\section{Implementation details of the procedures}\label{sec:supp-implementation}

This section collects the conventions that make REG-HD-TFKM and REG-HD-TRKM fully deterministic. They do not change the logic of the procedures described in Section~\ref{sec:method}, but they are needed to reproduce the reported results exactly. All numerical constants are fixed in advance and identical in every reported analysis.

\subsection{Fixed-parameter fit and multistart pool}\label{sec:supp-fit}

\paragraph{Empty groups and ties.}
If a retained group becomes empty, it is repaired before the scatter matrices are formed by reassigning high-norm retained observations, one per missing group, in the current retained-centered ambient coordinates. Nearest-centroid ties use MATLAB's first-minimum convention, and trimming uses \texttt{maxk} with no further rule at an exactly tied cutoff.

\paragraph{Validity of a start.}
A start is discarded only when the requested fit cannot be represented numerically: after an execution error, with too few retained observations, with fewer than $r$ finite generalized-eigen directions, or when finiteness or dimension checks fail. Reaching the 50-iteration cap is recorded separately: it means that an unchanged state was not certified, not that the terminal fit is unusable.

\paragraph{Seeds and initializers.}
Within a family, starts differ through seeded initial partitions. The seed of a start is the candidate seed plus $1000$ times the family index plus the within-family start index. Each trimmed K-means initializer runs for at most 20 iterations. Conditional on the data and the candidate seed, the whole pool of 100 starts is therefore reproducible.

\subsection{Choice of one solution per candidate}\label{sec:supp-choice}

\paragraph{HIER4 (TFKM).}
In both the within-family and the cross-family comparison, scores within $10^{-12}$ are treated as tied and resolved first by the smaller $J_{\rm trim}$, then by the first pool index. If only one family has valid fits, its within-family representative and within-family score are retained, and the cross-family components $A_{r,g,\alpha}$ and $J_{r,g,\alpha}$ are recorded as zero. Because self-comparisons are zero but kept in the denominator, the cross-family score is at most $1.2(F-1)/F$ for $F\ge2$ families with valid fits, so its range reflects the amount of cross-family evidence available.

\paragraph{RV--CV (TRKM).}
A fold is omitted if it contains fewer than two retained test observations or at most $g$ retained training observations. If an otherwise available training fold contains no member of one candidate group, its agreement is set to zero. A scoring view is invalid when it has at most $g$ retained observations or fewer than $g$ retained groups. Nonfinite view scores are omitted; if only one view remains finite, that view alone is used. Ties within $10^{-12}$ in the maximal RV--CV score are broken first by the larger separation between the candidate's groups in the two robust views and then by the smaller $J_{\rm trim}$; if no maximal-score candidate is available, the fit with the smallest objective is used as a fallback.

\subsection{Radial diagnostic}\label{sec:supp-radial}

Let $m_j$ and $\widetilde s_j$ be the coordinatewise median and MAD of the original variables, and let $\mathcal J_+$ contain the coordinates with positive finite MAD. Zero or nonfinite MADs are replaced by the floor
\[
s_{\rm floor}=\begin{cases}
\max\{\varepsilon_s,10^{-12}\operatorname{med}_{j\in\mathcal J_+}\widetilde s_j\},&\mathcal J_+\ne\varnothing,\\
1,&\mathcal J_+=\varnothing,
\end{cases}
\]
and $s_j=\max(\widetilde s_j,s_{\rm floor})$ after the corresponding finite-value convention. At least 20 finite log-radial scores are required. The one- and two-component Gaussian mixtures are fitted deterministically.

\subsection{Backward simplification: ranking keys}\label{sec:supp-backward}

\paragraph{TFKM.}
For a family representative $i$ at either state of a move, let $d_i$ count the opposite-state fits with mutually retained ARI at least $0.90$. Let $S_i$ be the percentile rank of this support among the endpoint representatives, $N_i$ the percentile rank of cross-family HIER4 evidence, $F_i$ the percentile rank of the fitted objective (lower is better) and $W_i$ the within-family HIER4 stability percentile. Endpoint fits are ranked lexicographically, in decreasing order, by
\[
\left(\min(S_i,N_i),\;\frac{S_i+N_i}{2},\;F_i,\;W_i,\;-J_{{\rm trim},i},\;-i\right).
\]
The fitted objective therefore enters only after cross-state support and stability. If different starting states end at different structural pairs, the result is reported as ambiguous rather than forcibly ranked.

\paragraph{TRKM.}
The starting states are defined in Section~\ref{sec:method-backward}. At each move, the pool at the current state and the pool at its predecessor are each restricted to the initialization family of their RV--CV-selected fit. A pair $(i,j)$ of fits from the two restricted pools is \emph{coherent} if their ARI on mutually retained observations is at least $0.90$. For fit $i$, let $Q_i^{RV}$ be the percentile rank of its RV--CV score within its pool and $Q_i^S$ the percentile rank of its number of coherent partners in the other pool. On the first move, coherent pairs are ranked lexicographically by
\[
\min(Q_i^S,Q_j^S),\quad
\min(Q_i^{RV},Q_j^{RV}),\quad
\frac{Q_i^{RV}+Q_j^{RV}}2,\quad
C_{ij},
\]
followed by deterministic index ties, where $C_{ij}$ is the ARI of the pair on mutually retained observations. If no coherent pair exists, the move is rejected and the path stops. After an accepted move the predecessor fit is carried forward; among its coherent partners at the next state, the choice uses the support percentile first, then the RV--CV percentile, then $C_{ij}$ and deterministic ties.

\section{Truth-assisted simulation diagnostics}\label{sec:supp-sim}

The quantities in this section use the generating labels, contamination indicators and subspace. They are never supplied to REG-HD-TFKM or REG-HD-TRKM and serve only to interpret the simulation results.

\subsection{Reference quantities and seeds}\label{sec:supp-reference}

Data seeds are $314159265+100s+b$, where $s$ is the scenario identifier (51001--51027 for MAIN, 51101--51105 for WEAK, 51228--51236 for STRUCTURE) and $b\in\{1,\ldots,30\}$ the replication. The candidate seed is the data seed plus $110000$ for REG-HD-TFKM and plus $220000$ for REG-HD-TRKM, and is the same for all 150 candidates of a data set. In the known-parameter benchmark the REG-HD fits are the stored candidates at $(r_0,g_0,\alpha_0)$; TKM, PCA--TKM, SSPCA--TKM, TCLUST and ROBPCA--TKM use the data seed plus $500000+10000k$ with $k=3,4,5,6,7$, respectively. If $n_k$ is the number of genuine observations with generating label $k$ and $h_0=\sum_k n_k$, define
\[
 \widehat\pi_k=n_k/h_0,\qquad
 a_k^{\rm ref}=\frac1{n_k}\sum_{i\in I_0:Y_i=k}U_0'X_i,\qquad
 \bar a^{\rm ref}=\sum_k\widehat\pi_k a_k^{\rm ref},
\]
\[
 B_a^{\rm ref}=\sum_k\widehat\pi_k
 (a_k^{\rm ref}-\bar a^{\rm ref})(a_k^{\rm ref}-\bar a^{\rm ref})',\qquad
 B_{\rm ref}=U_0B_a^{\rm ref}U_0',
\]
and $\beta_{\min}^{\rm ref}=\lambda_{r_0}(B_a^{\rm ref})$. The projected reference centroids are therefore sample means of the true-label genuine groups in the generating subspace, not population centroids. The model scatter matrices at the generating state are $\Sigma_0=U_0U_0'+\rho(I_p-U_0U_0')$ (the generating noise covariance), $\Sigma_B=B_{\rm ref}$ and $\Sigma_T=\Sigma_0+\Sigma_B$.

\paragraph{Spectral margin $G_{\rm spec}$.}
Let $h$ be the number of retained observations at the generating state, $S_W$, $S_B$ and $S_T$ the sample scatter matrices of Section~\ref{sec:method-ridge} at that state, and $\ell_W=\lambda_W/h$, $\ell_T=\lambda_T/h$ the normalized ridges. The subspace of each method is the leading invariant subspace of a whitened operator $K=D^{-1/2}AD^{-1/2}$, with $D=S_T/h+\ell_TI_p$ and $A=S_W/h+\ell_WI_p$ for TFKM (smallest eigenvalues) or $A=S_B/h$ for TRKM (largest eigenvalues). We write $\widehat K_M$ for this sample operator and $K_M^{\rm ref}$ for the reference operator obtained by replacing $S_W/h$, $S_B/h$ and $S_T/h$ with $\Sigma_0$, $\Sigma_B$ and $\Sigma_T$ and keeping the same $\ell_W$ and $\ell_T$. With $\delta_M^{\rm ref}$ the gap between the $r_0$th and $(r_0+1)$st eigenvalues of $K_M^{\rm ref}$, in the order used by the method, and $e_M^{\rm ref}=\|\widehat K_M-K_M^{\rm ref}\|_{\op}$, we set
\[
G_{\rm spec}=\delta_M^{\rm ref}/e_M^{\rm ref}.
\]
By the Davis--Kahan theorem, values well above one indicate that the estimated whitened subspace is close to the reference one.

\paragraph{Projected margin $G_{P,\mathrm{eff}}$.}
Let $\widehat U_M$ be the sample spectral subspace at the generating state, $c_\star=\lambda_{\min}(U_0'\widehat U_M\widehat U_M'U_0)$ its overlap with $U_0$, $d_{\min}$ the minimum distance between reference centroids, and $v_\rho(c)=1$ for $\rho\le1$ and $v_\rho(c)=c+\rho(1-c)$ for $\rho>1$. Then
\[
G_{P,\mathrm{eff}}=\frac{d_{\min}^2}{8\log n}\,\frac{c_\star}{v_\rho(c_\star)}.
\]
The factor $c_\star/v_\rho(c_\star)$ is a lower bound on the projected signal-to-noise ratio of every centroid contrast under the Gaussian model, and $G_{P,\mathrm{eff}}>1$ corresponds to a union-bound condition for exact nearest-centroid classification of the genuine observations. The margin is deliberately conservative: it is sufficient, not necessary.

\paragraph{Projected-oracle classification.}
Genuine observations are assigned to the nearest reference centroid after projection on $\widehat U_M$. Because $\widehat U_M$ is estimated from the same sample and the centroids use the true labels, this is an in-sample, truth-assisted calculation. It is exact in every WEAK data set for both methods, and in 1230/1230 TRKM and 1228/1230 TFKM data sets overall. Within WEAK, $G_{\rm spec}$ has AUC $0.9004$ for the availability of a high-ARI TRKM start and $0.6012$ for TFKM; the corresponding AUCs of $c_\star$, the mean overlap and $G_{P,\mathrm{eff}}$ are below $0.50$ for both methods.

\subsection{Group-number change point}\label{sec:supp-gpath}

For each data set, the distortion path is computed in the generating signal space on the genuine observations. For $g\le g_0$, all surjective assignments of the true-label groups to $g$ blocks are enumerated and the smallest within-block sum of squares is retained. For the first overfit gain, each true-label group is split by two-center Lloyd iterations from a leading-PC sign start, a farthest-pair start and eight seeded random-pair starts, with at most 100 iterations; the largest reduction found across groups and starts is used. Underfitting is thus represented by merges of whole components and the first overfit step by the best split found from these starts. Let $\widehat\Delta_{g+1}=\widehat R_g-\widehat R_{g+1}$ and
\[
 \widehat G_g^{\rm emp}
 =\frac{\min_{g<g_0}\widehat\Delta_{g+1}}
             {\widehat\Delta_{g_0+1}},
\]
with the denominator floored at machine epsilon. The separation flag requires the smallest pre-$g_0$ gain to exceed the first overfit gain by more than $10^{-12}$. A value above one establishes an empirical separating interval for this computed path; it does not provide a threshold chosen without $g_0$.

\begin{table}[htbp]
\centering
\caption{Empirical group-number change point. ``Separated'' is the fraction of data sets in which every pre-$g_0$ gain exceeds the first post-$g_0$ gain.}
\label{tab:supp-gpath}
\small
\begin{tabular}{llrrr}
\toprule
Panel & Contamination / $g_0$ & $N$ & Separated & Median $\widehat G_g^{\rm emp}$\\
\midrule
MAIN & bridge / 4    & 270 & 1.000 & 40.78\\
MAIN & clean / 4     & 270 & 1.000 & 42.45\\
MAIN & leverage / 4  & 270 & 1.000 & 40.78\\
STRUCTURE & bridge / 3 & 30 & 1.000 & 52.00\\
STRUCTURE & bridge / 5 & 60 & 1.000 & 25.14\\
STRUCTURE & clean / 3  & 30 & 1.000 & 51.40\\
STRUCTURE & clean / 5  & 60 & 1.000 & 26.14\\
STRUCTURE & leverage / 3 & 30 & 1.000 & 51.43\\
STRUCTURE & leverage / 5 & 60 & 1.000 & 24.60\\
WEAK & leverage / 4  & 150 & 1.000 & 19.33\\
\bottomrule
\end{tabular}
\end{table}

The first-crossing separation holds in all 1230 data sets (Table~\ref{tab:supp-gpath}); even the weakest group has a lower-decile ratio above approximately $15.9$. The generated samples therefore contain a clear group-number change point. This is evidence about the samples, not a replacement for the data-driven selector.

\subsection{Cross-fitted rank diagnostic}\label{sec:supp-rank}

Each true-label genuine group is randomly split into two disjoint halves with a fixed seed, with $\lfloor n_k/2\rfloor$ observations in the first half. The symmetrized cross-scatter $\widehat B_\times$ is formed from the ambient class means in the two halves, using the full genuine-sample weights $\widehat\pi_k$, and is compared with $B_{\rm ref}$. Set
\[
 t_{\rm diag}=\frac{\beta_{\min}^{\rm ref}}2,\qquad
 e_\times^{\rm ref}=\|\widehat B_\times-B_{\rm ref}\|_{\op},\qquad
 G_r^\times=\frac{\beta_{\min}^{\rm ref}}{2e_\times^{\rm ref}},
\]
with the denominator of $G_r^\times$ floored at machine epsilon. The Weyl flag is $e_\times^{\rm ref}<\beta_{\min}^{\rm ref}/2$, and $\widehat r_\times$ counts eigenvalues (not singular values, since the cross-scatter need not be positive semidefinite) greater than $t_{\rm diag}+10^{-12}$. Although the cross-scatter uses separate halves, the reference and threshold use true-group information from the full genuine sample; the calculation asks only whether the rank signal is visible once the grouping is known.

\begin{table}[htbp]
\centering
\caption{Cross-fitted rank diagnostic with an empirical true-label reference. ``Weyl pass'' is the sufficient event $e_\times^{\rm ref}<\beta_{\min}^{\rm ref}/2$; the last column is the direct midpoint rank calculation.}
\label{tab:supp-rank}
\small
\begin{tabular}{rrrrrr}
\toprule
$r_0$ & $g_0$ & $N$ & Median $G_r^\times$ & Weyl pass & Exact $\widehat r_\times$\\
\midrule
2 & 3 & 90  & 1.574 & 1.000 & 1.000\\
3 & 4 & 960 & 1.219 & 0.589 & 1.000\\
3 & 5 & 90  & 1.314 & 1.000 & 1.000\\
4 & 5 & 90  & 1.019 & 0.567 & 1.000\\
\bottomrule
\end{tabular}
\end{table}

The midpoint diagnostic recovers $r_0$ in all 1230 data sets (Table~\ref{tab:supp-rank}), including every $(3,4)$ and $(4,5)$ data set in which the sufficient Weyl event fails. The direct rank signal is therefore stronger than the conservative perturbation flag suggests.

\subsection{Sensitivity of the trimming decision to its thresholds}\label{sec:supp-thresholds}

The thresholds of the trimming rules were fixed before the simulations reported here and were not varied there. Because the statistics on which these rules act are stored for every data set, the decisions that other thresholds would have produced can be recomputed without refitting. Table~\ref{tab:supp-thresholds} reports, for each rule, the number of positive decisions under alternative thresholds, separately for the 510 leverage data sets (where trimming is the target) and the 720 clean and bridge data sets (where it is not).

\begin{table}[htbp]
\centering
\caption{Trimming decisions under alternative thresholds, recomputed from the stored statistics of the simulation study. Entries are numbers of data sets with a positive decision. Default values are in bold.}
\label{tab:supp-thresholds}
\footnotesize
\setlength{\tabcolsep}{3pt}
\begin{tabular}{llrr}
\toprule
Rule & Thresholds & Leverage (510) & Clean/bridge (720)\\
\midrule
Radial & $\Delta BIC>\mathbf{0}$, $w_H\le 0.15$ to $\mathbf{0.30}$ & 510 & 2\\
       & $\Delta BIC>0$, $w_H\le 0.40$ / $0.50$ & 510 & 3 / 5\\
       & $\Delta BIC>5$, $w_H\le 0.15$ to $0.40$ & 510 & 0\\
       & $\Delta BIC>c$, $c\in[10,300]$, $w_H\le 0.15$ to $0.50$ & 510 & 0\\
\addlinespace
TFKM path & $J_{\max}\ge\mathbf{0.50}$, $D\ge 0.07$ / $\mathbf{0.12}$ / $0.17$ & -- & 16 / 4 / 1\\
          & $J_{\max}\ge 0.55$, $D\ge 0.07$ / $0.12$ / $0.17$ & -- & 4 / 1 / 0\\
          & $J_{\max}\ge 0.45$, $D\ge 0.07$ / $0.12$ / $0.17$ & -- & 29 / 5 / 1\\
\addlinespace
TRKM path & $\Delta RV\ge\mathbf{0.025}$ / $0.030$ / $0.035$ / $0.040$ & -- & 21 / 5 / 1 / 0\\
\bottomrule
\end{tabular}
\end{table}

For the radial rule, the two populations are far apart. The BIC contrast is at least 314 in every leverage data set and at most 16.6 in every clean or bridge data set, and the estimated outer weight equals the true fraction $0.10$ in every leverage data set. Every cutoff for $\Delta BIC$ between 0 and 300 and every bound on $w_H$ between $0.15$ and $0.50$ therefore detects all 510 leverage data sets, with at most 5 positives among the 720 others. For the TFKM path rule, the median cutoff for $A_{\rm med}$ has no effect over $[0.34,0.54]$ at the default $J_{\max}$ and $D$ cutoffs, because the rule is driven by $J_{\max}$ and $D$; with $J_{\max}\ge 0.50$ the number of positives among clean and bridge data sets stays between 1 and 16 (at most 2.2\%) for $D$ cutoffs between $0.07$ and $0.17$, whereas it rises to 29 when the $J_{\max}$ cutoff is lowered to $0.45$. Because leverage is detected by the radial branch, these changes affect only the clean and bridge data sets. In the log-level application, the positive TFKM decision rests on $J_{\max}=0.512$ and would not survive a cutoff of $0.55$. For the TRKM path rule, the 21 positives among clean and bridge data sets all have $\Delta RV$ between $0.025$ and $0.0375$; a cutoff of $0.04$ would remove them, whereas the positive decision in the ratio application ($\Delta RV=0.059$) is unaffected by any cutoff up to $0.05$. Only increases of this cutoff can be evaluated from the stored statistics, because data sets with $\Delta RV<0.025$ were not tested for the onset condition. Thresholds that act on the structure (the $0.90$ ARI requirement, the $0.01$ reconstruction tolerance and the weight $0.2$) cannot be varied without refitting and are not covered by this analysis.

\subsection{Results by design cell}\label{sec:supp-cells}

Table~\ref{tab:supp-cells} reports the three measures of Table~\ref{tab:primary-performance} for each of the 41 design cells; every cell has 30 data sets.

\begin{table}[htbp]
\centering
\caption{Automatic recovery by design cell (30 data sets per cell). Columns as in Table~\ref{tab:primary-performance}.}
\label{tab:supp-cells}
\footnotesize
\setlength{\tabcolsep}{3.5pt}
\begin{tabular}{lrrcrl ccc ccc}
\toprule
 & & & & & & \multicolumn{3}{c}{REG-HD-TFKM} & \multicolumn{3}{c}{REG-HD-TRKM}\\
\cmidrule(lr){7-9}\cmidrule(lr){10-12}
Panel & $p/n$ & $\rho$ & $(r_0,g_0)$ & $\delta$ & Contamination & $(r,g)$ & Trim & Joint & $(r,g)$ & Trim & Joint\\
\midrule
MAIN & 0.5 & 0.25 & $(3,4)$ & 10 & Clean & 1.000 & 1.000 & 1.000 & 0.933 & 1.000 & 0.933\\
 & 0.5 & 0.25 & $(3,4)$ & 10 & Bridge & 0.900 & 0.967 & 0.867 & 0.967 & 0.967 & 0.933\\
 & 0.5 & 0.25 & $(3,4)$ & 10 & Leverage & 0.967 & 1.000 & 0.967 & 0.967 & 1.000 & 0.967\\
 & 0.5 & 1 & $(3,4)$ & 10 & Clean & 0.767 & 1.000 & 0.767 & 0.933 & 1.000 & 0.933\\
 & 0.5 & 1 & $(3,4)$ & 10 & Bridge & 0.933 & 1.000 & 0.933 & 0.833 & 0.867 & 0.700\\
 & 0.5 & 1 & $(3,4)$ & 10 & Leverage & 0.967 & 1.000 & 0.967 & 0.933 & 1.000 & 0.933\\
 & 0.5 & 2 & $(3,4)$ & 10 & Clean & 0.933 & 0.967 & 0.900 & 0.967 & 0.967 & 0.967\\
 & 0.5 & 2 & $(3,4)$ & 10 & Bridge & 0.933 & 1.000 & 0.933 & 0.933 & 1.000 & 0.933\\
 & 0.5 & 2 & $(3,4)$ & 10 & Leverage & 0.767 & 1.000 & 0.767 & 0.733 & 1.000 & 0.733\\
 & 1 & 0.25 & $(3,4)$ & 10 & Clean & 0.867 & 1.000 & 0.867 & 0.933 & 1.000 & 0.933\\
 & 1 & 0.25 & $(3,4)$ & 10 & Bridge & 0.867 & 0.967 & 0.833 & 0.900 & 0.967 & 0.867\\
 & 1 & 0.25 & $(3,4)$ & 10 & Leverage & 1.000 & 1.000 & 1.000 & 0.567 & 1.000 & 0.567\\
 & 1 & 1 & $(3,4)$ & 10 & Clean & 0.833 & 0.967 & 0.833 & 0.967 & 1.000 & 0.967\\
 & 1 & 1 & $(3,4)$ & 10 & Bridge & 0.833 & 1.000 & 0.833 & 0.967 & 1.000 & 0.967\\
 & 1 & 1 & $(3,4)$ & 10 & Leverage & 0.767 & 1.000 & 0.767 & 0.600 & 1.000 & 0.600\\
 & 1 & 2 & $(3,4)$ & 10 & Clean & 0.767 & 1.000 & 0.767 & 0.867 & 1.000 & 0.867\\
 & 1 & 2 & $(3,4)$ & 10 & Bridge & 0.900 & 1.000 & 0.900 & 1.000 & 1.000 & 1.000\\
 & 1 & 2 & $(3,4)$ & 10 & Leverage & 0.567 & 1.000 & 0.567 & 0.600 & 1.000 & 0.600\\
 & 2 & 0.25 & $(3,4)$ & 10 & Clean & 0.833 & 1.000 & 0.833 & 1.000 & 1.000 & 1.000\\
 & 2 & 0.25 & $(3,4)$ & 10 & Bridge & 0.900 & 0.967 & 0.867 & 1.000 & 0.967 & 0.967\\
 & 2 & 0.25 & $(3,4)$ & 10 & Leverage & 1.000 & 1.000 & 1.000 & 0.700 & 1.000 & 0.700\\
 & 2 & 1 & $(3,4)$ & 10 & Clean & 0.900 & 1.000 & 0.900 & 0.800 & 1.000 & 0.800\\
 & 2 & 1 & $(3,4)$ & 10 & Bridge & 0.867 & 1.000 & 0.867 & 0.933 & 1.000 & 0.933\\
 & 2 & 1 & $(3,4)$ & 10 & Leverage & 0.633 & 1.000 & 0.633 & 0.567 & 1.000 & 0.567\\
 & 2 & 2 & $(3,4)$ & 10 & Clean & 0.867 & 1.000 & 0.867 & 0.933 & 1.000 & 0.933\\
 & 2 & 2 & $(3,4)$ & 10 & Bridge & 0.733 & 1.000 & 0.733 & 0.767 & 0.833 & 0.633\\
 & 2 & 2 & $(3,4)$ & 10 & Leverage & 0.467 & 1.000 & 0.467 & 0.500 & 1.000 & 0.500\\
\addlinespace
STRUCTURE & 1 & 1 & $(2,3)$ & 10 & Clean & 0.900 & 1.000 & 0.900 & 1.000 & 1.000 & 1.000\\
 & 1 & 1 & $(2,3)$ & 10 & Bridge & 0.833 & 1.000 & 0.833 & 0.933 & 1.000 & 0.933\\
 & 1 & 1 & $(2,3)$ & 10 & Leverage & 0.967 & 1.000 & 0.967 & 0.567 & 1.000 & 0.567\\
 & 1 & 1 & $(3,5)$ & 10 & Clean & 0.367 & 1.000 & 0.367 & 0.000 & 0.867 & 0.000\\
 & 1 & 1 & $(3,5)$ & 10 & Bridge & 0.433 & 0.967 & 0.433 & 0.133 & 0.867 & 0.067\\
 & 1 & 1 & $(3,5)$ & 10 & Leverage & 0.067 & 1.000 & 0.067 & 0.200 & 1.000 & 0.200\\
 & 1 & 1 & $(4,5)$ & 10 & Clean & 0.767 & 1.000 & 0.767 & 0.833 & 1.000 & 0.833\\
 & 1 & 1 & $(4,5)$ & 10 & Bridge & 0.933 & 1.000 & 0.933 & 0.900 & 0.933 & 0.833\\
 & 1 & 1 & $(4,5)$ & 10 & Leverage & 0.867 & 1.000 & 0.867 & 0.200 & 1.000 & 0.200\\
\addlinespace
WEAK & 1 & 1 & $(3,4)$ & 7 & Leverage & 0.567 & 1.000 & 0.567 & 0.400 & 1.000 & 0.400\\
 & 1 & 2 & $(3,4)$ & 7 & Leverage & 0.300 & 1.000 & 0.300 & 0.267 & 1.000 & 0.267\\
 & 2 & 0.25 & $(3,4)$ & 7 & Leverage & 0.833 & 1.000 & 0.833 & 0.500 & 1.000 & 0.500\\
 & 2 & 1 & $(3,4)$ & 7 & Leverage & 0.333 & 1.000 & 0.333 & 0.600 & 1.000 & 0.600\\
 & 2 & 2 & $(3,4)$ & 7 & Leverage & 0.067 & 1.000 & 0.067 & 0.033 & 1.000 & 0.033\\
\bottomrule
\end{tabular}
\end{table}

\clearpage
\end{appendices}


\begin{thebibliography}{42}
\providecommand{\natexlab}[1]{#1}
\providecommand{\url}[1]{{#1}}
\providecommand{\urlprefix}{URL }
\providecommand{\doi}[1]{\url{https://doi.org/#1}}
\providecommand{\eprint}[2][]{\url{#2}}
 \bibcommenthead

\bibitem[{Aggarwal et~al.(1999)Aggarwal, Procopiuc, Wolf, Yu, and
  Park}]{aggarwal1999projected}
Aggarwal CC, Procopiuc C, Wolf JL, et~al (1999) Fast algorithms for projected
  clustering. In: Proceedings of the 1999 ACM SIGMOD International Conference
  on Management of Data, pp 61--72, \doi{10.1145/304181.304188}

\bibitem[{{\'A}lvarez-Esteban and
  Garc{\'i}a-Escudero(2022)}]{alvarez2022robust}
{\'A}lvarez-Esteban PC, Garc{\'i}a-Escudero LA (2022) Robust clustering of
  functional directional data. Advances in Data Analysis and Classification
  16:181--199. \doi{10.1007/s11634-021-00482-3}

\bibitem[{Arabie and Hubert(1994)}]{arabie1994cluster}
Arabie P, Hubert L (1994) Cluster analysis in marketing research. In: Bagozzi
  RP (ed) Advanced Methods of Marketing Research. Blackwell, Cambridge, MA, p
  160--189

\bibitem[{Bouveyron and Brunet(2012)}]{bouveyron2012fisher}
Bouveyron C, Brunet C (2012) Simultaneous model-based clustering and
  visualization in the fisher discriminative subspace. Statistics and Computing
  22(1):301--324. \doi{10.1007/s11222-011-9249-9}

\bibitem[{Bouveyron and Brunet-Saumard(2014)}]{bouveyron2014model}
Bouveyron C, Brunet-Saumard C (2014) Model-based clustering of high-dimensional
  data: A review. Computational Statistics \& Data Analysis 71:52--78.
  \doi{10.1016/j.csda.2012.12.008}

\bibitem[{Bouveyron et~al.(2007)Bouveyron, Girard, and
  Schmid}]{bouveyron2007high}
Bouveyron C, Girard S, Schmid C (2007) High-dimensional data clustering.
  Computational Statistics \& Data Analysis 52(1):502--519.
  \doi{10.1016/j.csda.2007.02.009}

\bibitem[{Brodinov{\'a} et~al.(2019)Brodinov{\'a}, Filzmoser, Ortner,
  Breiteneder, and Rohm}]{brodinova2019robust}
Brodinov{\'a} {\v S}, Filzmoser P, Ortner T, et~al (2019) Robust and sparse
  k-means clustering for high-dimensional data. Advances in Data Analysis and
  Classification 13:905--932. \doi{10.1007/s11634-019-00356-9}

\bibitem[{Coraggio and Coretto(2023)}]{coraggio2023selecting}
Coraggio L, Coretto P (2023) Selecting the number of clusters, clustering
  models, and algorithms: A unifying approach based on the quadratic
  discriminant score. Journal of Multivariate Analysis 196:105181.
  \doi{10.1016/j.jmva.2023.105181}

\bibitem[{Cuesta-Albertos et~al.(1997)Cuesta-Albertos, Gordaliza, and
  Matr{\'a}n}]{cuesta1997trimmed}
Cuesta-Albertos JA, Gordaliza A, Matr{\'a}n C (1997) Trimmed k-means: An
  attempt to robustify quantizers. The Annals of Statistics 25(2):553--576.
  \doi{10.1214/aos/1031833664}

\bibitem[{De~Soete and Carroll(1994)}]{desoete1994kmeans}
De~Soete G, Carroll JD (1994) K-means clustering in a low-dimensional euclidean
  space. In: Diday E, Lechevallier Y, Schader M, et~al (eds) New Approaches in
  Classification and Data Analysis. Springer, Berlin, p 212--219,
  \doi{10.1007/978-3-642-51175-2_24}

\bibitem[{Elhamifar and Vidal(2013)}]{elhamifar2013sparse}
Elhamifar E, Vidal R (2013) Sparse subspace clustering: Algorithm, theory, and
  applications. IEEE Transactions on Pattern Analysis and Machine Intelligence
  35(11):2765--2781. \doi{10.1109/TPAMI.2013.57}

\bibitem[{Farn{\`e} and Camillo(2025)}]{farne2025trimmed}
Farn{\`e} M, Camillo F (2025) Trimmed factorial k-means: A clustering
  application to a cookie dataset. In: Statistical Models and Learning Methods
  for Complex Data. Springer, p 103--111, \doi{10.1007/978-3-031-84702-8_12}

\bibitem[{Farn{\`e} and Vouldis(2021)}]{farne2021banks}
Farn{\`e} M, Vouldis AT (2021) Banks' business models in the euro area: a
  cluster analysis in high dimensions. Annals of Operations Research
  305(1--2):23--57. \doi{10.1007/s10479-021-04045-9}

\bibitem[{Fritz et~al.(2012)Fritz, Garc{\'i}a-Escudero, and
  Mayo-Iscar}]{fritz2012tclust}
Fritz H, Garc{\'i}a-Escudero LA, Mayo-Iscar A (2012) {tclust}: An {R} package
  for a trimming approach to cluster analysis. Journal of Statistical Software
  47(12):1--26. \doi{10.18637/jss.v047.i12}

\bibitem[{Garc{\'i}a-Escudero and Mayo-Iscar(2024)}]{garciaescudero2024review}
Garc{\'i}a-Escudero LA, Mayo-Iscar A (2024) Robust clustering based on
  trimming. WIREs Computational Statistics 16(4):e1658. \doi{10.1002/wics.1658}

\bibitem[{Garc{\'i}a-Escudero et~al.(2008)Garc{\'i}a-Escudero, Gordaliza,
  Matr{\'a}n, and Mayo-Iscar}]{garciaescudero2008general}
Garc{\'i}a-Escudero LA, Gordaliza A, Matr{\'a}n C, et~al (2008) A general
  trimming approach to robust cluster analysis. The Annals of Statistics
  36(3):1324--1345. \doi{10.1214/07-AOS515}

\bibitem[{Garc{\'i}a-Escudero et~al.(2026)Garc{\'i}a-Escudero, Hennig,
  Mayo-Iscar, Morelli, and Riani}]{garciaescudero2026choice}
Garc{\'i}a-Escudero LA, Hennig C, Mayo-Iscar A, et~al (2026) Choice of trimming
  proportion and number of clusters in robust clustering based on trimming.
  Journal of Computational and Graphical Statistics 35(2):606--618.
  \doi{10.1080/10618600.2025.2554675}

\bibitem[{Gattone and Rocci(2012)}]{gattone2012clustering}
Gattone SA, Rocci R (2012) Clustering curves on a reduced subspace. Journal of
  Computational and Graphical Statistics 21(2):361--379.
  \doi{10.1080/10618600.2012.679237}

\bibitem[{Golub and Van~Loan(2013)}]{golub2013matrix}
Golub GH, Van~Loan CF (2013) Matrix Computations, 4th edn. Johns Hopkins
  University Press, Baltimore

\bibitem[{Greco et~al.(2020)Greco, Lucadamo, and Amenta}]{greco2020impartial}
Greco L, Lucadamo A, Amenta P (2020) An impartial trimming approach for joint
  dimension and sample reduction. Journal of Classification 37(3):769--788.
  \doi{10.1007/s00357-019-09354-0}

\bibitem[{Hubert and Arabie(1985)}]{hubert1985comparing}
Hubert L, Arabie P (1985) Comparing partitions. Journal of Classification
  2(1):193--218. \doi{10.1007/BF01908075}

\bibitem[{Hubert et~al.(2005)Hubert, Rousseeuw, and
  Vanden~Branden}]{hubert2005robpca}
Hubert M, Rousseeuw PJ, Vanden~Branden K (2005) {ROBPCA}: A new approach to
  robust principal component analysis. Technometrics 47(1):64--79.
  \doi{10.1198/004017004000000563}

\bibitem[{Kriegel et~al.(2009)Kriegel, Kr{\"o}ger, and
  Zimek}]{kriegel2009survey}
Kriegel HP, Kr{\"o}ger P, Zimek A (2009) Clustering high-dimensional data: A
  survey on subspace clustering, pattern-based clustering, and correlation
  clustering. ACM Transactions on Knowledge Discovery from Data 3(1):1--58.
  \doi{10.1145/1497577.1497578}

\bibitem[{Liu et~al.(2013)Liu, Lin, Yan, Sun, Yu, and Ma}]{liu2013robust}
Liu G, Lin Z, Yan S, et~al (2013) Robust recovery of subspace structures by
  low-rank representation. IEEE Transactions on Pattern Analysis and Machine
  Intelligence 35(1):171--184. \doi{10.1109/TPAMI.2012.88}

\bibitem[{Locantore et~al.(1999)Locantore, Marron, Simpson, Tripoli, Zhang, and
  Cohen}]{locantore1999robust}
Locantore N, Marron JS, Simpson DG, et~al (1999) Robust principal component
  analysis for functional data. Test 8(1):1--73. \doi{10.1007/BF02595862}

\bibitem[{Mori and Anderlucci(2026)}]{mori2026random}
Mori M, Anderlucci L (2026) Model-based clustering of functional data via
  random projection ensembles. Advances in Data Analysis and Classification
  20:877--901. \doi{10.1007/s11634-026-00684-7}

\bibitem[{Pollard(1981)}]{pollard1981strong}
Pollard D (1981) Strong consistency of k-means clustering. The Annals of
  Statistics 9(1):135--140. \doi{10.1214/aos/1176345339}

\bibitem[{Riani et~al.(2012)Riani, Perrotta, and Torti}]{riani2012fsda}
Riani M, Perrotta D, Torti F (2012) {FSDA}: A {MATLAB} toolbox for robust
  analysis and interactive data exploration. Chemometrics and Intelligent
  Laboratory Systems 116:17--32. \doi{10.1016/j.chemolab.2012.03.017}

\bibitem[{Rocci et~al.(2011)Rocci, Gattone, and Vichi}]{rocci2011new}
Rocci R, Gattone SA, Vichi M (2011) A new dimension reduction method: Factor
  discriminant k-means. Journal of Classification 28(2):210--226.
  \doi{10.1007/s00357-011-9085-9}

\bibitem[{Sugar and James(2003)}]{sugar2003clusters}
Sugar CA, James GM (2003) Finding the number of clusters in a dataset: An
  information-theoretic approach. Journal of the American Statistical
  Association 98(463):750--763. \doi{10.1198/016214503000000666}

\bibitem[{Terada(2014)}]{terada2014strong}
Terada Y (2014) Strong consistency of reduced k-means clustering. Scandinavian
  Journal of Statistics 41(4):913--931. \doi{10.1111/sjos.12074}

\bibitem[{Terada(2015)}]{terada2015strong}
Terada Y (2015) Strong consistency of factorial k-means clustering. Annals of
  the Institute of Statistical Mathematics 67(2):335--357.
  \doi{10.1007/s10463-014-0454-0}

\bibitem[{Tibshirani and Walther(2005)}]{tibshirani2005cluster}
Tibshirani R, Walther G (2005) Cluster validation by prediction strength.
  Journal of Computational and Graphical Statistics 14(3):511--528.
  \doi{10.1198/106186005X59243}

\bibitem[{Tibshirani et~al.(2001)Tibshirani, Walther, and
  Hastie}]{tibshirani2001gap}
Tibshirani R, Walther G, Hastie T (2001) Estimating the number of clusters in a
  data set via the gap statistic. Journal of the Royal Statistical Society:
  Series B (Statistical Methodology) 63(2):411--423.
  \doi{10.1111/1467-9868.00293}

\bibitem[{Timmerman et~al.(2010)Timmerman, Ceulemans, Kiers, and
  Vichi}]{timmerman2010factorial}
Timmerman ME, Ceulemans E, Kiers HAL, et~al (2010) Factorial and reduced
  k-means reconsidered. Computational Statistics \& Data Analysis
  54(7):1858--1871. \doi{10.1016/j.csda.2010.02.009}

\bibitem[{Torti et~al.(2019)Torti, Perrotta, Riani, and
  Cerioli}]{torti2019trimming}
Torti F, Perrotta D, Riani M, et~al (2019) Assessing trimming methodologies for
  clustering linear regression data. Advances in Data Analysis and
  Classification 13:227--257. \doi{10.1007/s11634-018-0331-4}

\bibitem[{Verboven and Hubert(2005)}]{verboven2005libra}
Verboven S, Hubert M (2005) {LIBRA}: a {MATLAB} library for robust analysis.
  Chemometrics and Intelligent Laboratory Systems 75(2):127--136.
  \doi{10.1016/j.chemolab.2004.06.003}

\bibitem[{Vichi and Kiers(2001)}]{vichi2001factorial}
Vichi M, Kiers HAL (2001) Factorial k-means analysis for two-way data.
  Computational Statistics \& Data Analysis 37(1):49--64.
  \doi{10.1016/S0167-9473(00)00064-5}

\bibitem[{Vichi and Saporta(2009)}]{vichi2009clustering}
Vichi M, Saporta G (2009) Clustering and disjoint principal component analysis.
  Computational Statistics \& Data Analysis 53(8):3194--3208.
  \doi{10.1016/j.csda.2008.05.028}

\bibitem[{Vidal(2011)}]{vidal2011subspace}
Vidal R (2011) Subspace clustering. IEEE Signal Processing Magazine
  28(2):52--68. \doi{10.1109/MSP.2010.939739}

\bibitem[{Witten and Tibshirani(2010)}]{witten2010framework}
Witten DM, Tibshirani R (2010) A framework for feature selection in clustering.
  Journal of the American Statistical Association 105(490):713--726.
  \doi{10.1198/jasa.2010.tm09415}

\bibitem[{Yang et~al.(2017)Yang, Xiang, and Yao}]{yang2017robust}
Yang L, Xiang S, Yao W (2017) Robust fitting of mixtures of factor analyzers
  using the trimmed likelihood estimator. Communications in
  Statistics---Simulation and Computation 46(2):1280--1291.
  \doi{10.1080/03610918.2014.999088}

\end{thebibliography}
\end{document}